\documentclass[a4paper,11pt]{article}

\usepackage{jheppub} 

\usepackage{axodraw4j}
\newcommand\bqa{\begin{eqnarray}}
\newcommand\eqa{\end{eqnarray}}
\newcommand\nl{\nonumber \\}

\newcommand\qot{q_{1\!2}}
\newcommand\qots{q^2_{1\!2}}
\newcommand\qotsb{\bar q^2_{1\!2}}

\newcommand\muuv{\mu_{\mbox{\tiny UV}}}
\newcommand\muir{\mu_{\mbox{\tiny IR}}}
\newcommand\mur{\mu_{\mbox{\tiny R}}}

\newcommand\toGP{\,~{\to}^{\vphantom{a}^{\hskip -11pt {\mbox{\tiny \rm GP}}}}~\,}
\newcommand\toSIC{\,~{\to}^{\vphantom{a}^{\hskip -11pt {\mbox{\tiny \rm SIC}}}}~\,}
\newcommand\toGS{\,~{\to}^{\vphantom{a}^{\hskip -11pt {\mbox{\tiny \rm GS}}}}~\,}
\newcommand\dkl{\delta_{kl}}

\title{\boldmath NNLO final-state quark-pair corrections in four dimensions}

\author[a]{B. Page,}
\author[b]{R. Pittau}

\affiliation[a]{Physikalisches Institut, Albert-Ludwigs-Universitat Freiburg, D79104 Freiburg, Germany}
\affiliation[b]{Departamento de F\'isica Te\'orica y del Cosmos and CAFPE, Universidad de Granada, Campus Fuentenueva s.n., E-18071 Granada, Spain}

\emailAdd{ben.page@physik.uni-freiburg.de}
\emailAdd{pittau@ugr.es}

\abstract{We describe how NNLO final state quark-pair corrections are computed in FDR by directly enforcing gauge invariance and unitarity in the definition of the regularized divergent integrals. The whole procedure is strictly four-dimensional and renormalization is performed by simply fixing bare parameters in terms of physical measurements. We give details of our approach and demonstrate how virtual and real contributions can be merged together without relying on dimensional regularization. As an example, we recompute the ${ H} \to b \bar b + {jets}$ and $\gamma^\ast \to {jets}$ inclusive rates at the NNLO accuracy in the large $N_F$ limit of QCD.}

\begin{document} 
\maketitle

\flushbottom

\section{Introduction}
\label{sec:intro}
It is well known that calculations of observables in quantum field theory (QFT) are complicated by divergences in intermediate steps. On general grounds these divergences fall into two categories - ultraviolet (UV) divergences, associated with short wavelengths, and infrared (IR) divergences, associated with large wavelengths and/or collinear configurations. They show up in the form of integrals which do not exist in the four-dimensional physical space.

The customary way to deal with UV infinities is a two step procedure. Firstly, one regularizes the divergent integrals. Secondly, one reabsorbs the UV pieces into the free parameters of the Lagrangian.
As for the IR infinities, they cancel when considering sufficiently inclusive observables, or can be reabsorbed in the parton densities.
Depending on the approach, this cancellation can be achieved before or after integration. In the latter case, IR divergent integrals also need to be regulated.

When performing calculations of observables in QFT, the exact method of regulating the UV/IR divergences is arbitrary. Nevertheless, this freedom is not absolute as the chosen method should not interfere with two core tenets of QFT, that are
\begin{subequations}\label{eq:prin}
  \bqa
  \label{eq:prin:1}  
  &\bullet&\mbox{gauge invariance;\hspace{10.5cm}} \\
  \label{eq:prin:2}  
  &\bullet&\mbox{{unitarity}.\hspace{10.5cm}}
  \eqa
\end{subequations}
These general principles have concrete consequences in perturbative calculations. Gauge invariance implies a set of graphical identities (see e.g. \cite{Sterman:1994ce}) that need to be fulfilled to all perturbative orders by the Feynman diagrams of the QFT.
Unitarity, meanwhile, demands the validity of the cutting equations
\cite{Cutkosky:1960sp,Veltman:1963th} corresponding to the relation
\bqa
\label{eq:unitarity}
i (T -T^{\dag}) = - T^\dag T
\eqa
for the $T$ matrix~\cite{tHooft:1973wag}.
Both \eqref{eq:prin} are essential to preserve the Ward/Slavnov-Taylor
identities (WI) at the regularized level, known as the
regularized quantum action principle \cite{Stockinger:2005gx}.

The most commonly used technique is dimensional regularization \cite{'tHooft:1972fi,Bollini:1972ui} (DReg). DReg exploits the fact that
gauge invariance and unitarity are naturally preserved as they hold for the theory in all values of the dimensionality $d$ of the space-time. Hence, the divergent integrals are analytically evaluated in $d$ dimensions, and the asymptotic $d \to 4$ limit is eventually taken. By doing so, UV/IR divergences are parametrized in terms of negative powers of a Laurent expansion in $(d-4)$.
In this framework, one still has some freedom to define objects in intermediate steps. Hence, several variants of DReg exist such as conventional dimensional regularization \cite{Collins:1984xc}, 'tHooft-Veltman \cite{'tHooft:1972fi}, four-dimensional helicity \cite{Bern:1991aq} and dimensional reduction \cite{Siegel:1979wq}. We refer to \cite{Signer:2008va} for an exact definition of all of them. 

In recent years, a considerable effort has been pursued by several groups to introduce more four-dimensional ingredients in the definition
of regularization. The main motivation being an attempt to simplify analytical and numerical methods, as well as to try to consider different theoretical perspectives.
This four-dimensional program has resulted in a number of methods such as the four-dimensional formulation of FDH \cite{Fazio:2014xea}, implicit regularization \cite{Battistel:1998sz,Cherchiglia:2010yd}, four-dimensional unsubtraction \cite{Hernandez-Pinto:2015ysa,Sborlini:2016gbr,Sborlini:2016hat} and FDR \cite{Pittau:2012zd,Donati:2013iya,Donati:2013voa,Pittau:2013qla,Page:2015zca}.
They are described and compared in \cite{Gnendiger:2017pys}.
Whilst all approaches are different, FDR is the only method that does not rely on the customary UV renormalization procedure. Instead, the result of an FDR calculation is directly a renormalized quantity.

FDR treats UV divergences by performing a subtraction, extracting from the loop integrands the divergent parts which do not contain physical information - the so-called {\em vacua}.
In the case of IR finite amplitudes, the validity of the FDR strategy has been explicitly demonstrated in \cite{Page:2015zca}.
In the presence of IR divergent configurations, the IR regulator should not interfere with principles \eqref{eq:prin}. If this is achieved, the correct physical result is obtained for IR safe quantities.  At NLO, it is known how to match real and one-loop contributions in the presence of final state IR singularities \cite{Pittau:2013qla,Gnendiger:2017pys}, while, so far, no FDR NNLO calculation has been performed involving infrared divergent configurations. In this paper we bridge this gap and give the first example of such a computation. We describe how NNLO final state quark-pair corrections can be computed in FDR in a way that automatically respects the principles \eqref{eq:prin}. In particular, we reproduce the
${\rm \overline{MS}}$ results for the $N_F$ part of
${ H} \to b \bar b + {jets}$ and $\gamma^\ast \to {jets}$. 
Since DReg is never used, explicitly or implicitly, this represents, to our knowledge, the first example of a realistic fully four-dimensional NNLO calculation.

The structure of the paper is as follows. In section \ref{sec:fdr} we present our definitions of the virtual and real integrals appearing when computing NNLO quark-pair corrections. In section \ref{sec:ren} we discuss the relevant renormalization issues.
Sections \ref{sec:H} and \ref{sec:gamma} give a detailed description
of the  ${ H} \to b \bar b + {jets}$ and $\gamma^\ast \to {jets}$
calculations. Finally, in section \ref{sec:conc} we summarize our findings and discuss perspectives and directions opened by the procedures introduced in this work.

\section{Inclusive quark-pair corrections and FDR}
\label{sec:fdr}
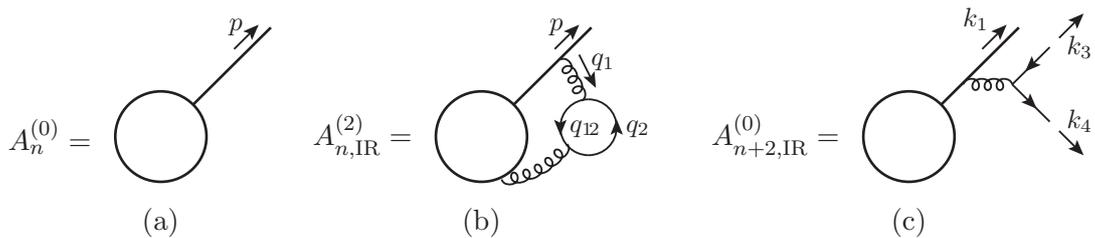
\begin{figure}[tbp]
\begin{picture}(215,90)(-220,-47)
\SetScale{1}
\SetWidth{1}
\SetOffset(-140,0)
\Text(-40,-14)[r]{$A^{(0)}_n =$}
\Line(-4,-4)(27,27)
\BCirc(-14,-14){17}
\Text(-14,-41)[t]{(a)}
\SetScale{0.7}
\LongArrow(19,27)(29,37)
\Text(14,24)[b]{\small $p$}
\SetScale{1}
\SetOffset(-20,0)
\Text(-40,-14)[r]{$A^{(2)}_{n,\rm IR} =$}
\Line(-4,-4)(27,27)
\BCirc(-14,-14){17}
\Text(-14,-41)[t]{(b)}
\SetScale{0.7}
\ArrowArc(37,-15)(15,-90,90)
\ArrowArc(37,-15)(15,90,270)
\GlueArc(-20,-14)(55,14,40){3}{3}
\GlueArc(-10,0)(43,-89,-34){3}{5}
\SetScale{0.7}
\LongArrow(19,27)(29,37)
\Text(14,24)[b]{\small $p$}
\LongArrow(32,24)(39,10)
\Text(27,14)[l]{\small $q_1$}
\Text(40,-11)[l]{\small $q_2$}
\Text(19,-11)[l]{\small $\qot$}
\SetScale{1}
\SetOffset(140,0)
\Text(-40,-14)[r]{$A^{(0)}_{n+2,\rm IR} =$}
\Line(-4,-4)(27,27)
\BCirc(-14,-14){17}
\Text(-14,-41)[t]{(c)}
\SetScale{0.7}
\Gluon(9,8)(35,8){3}{3}
\ArrowLine(55,28)(35,8)
\ArrowLine(35,8)(55,-12)
\SetScale{0.7}
\LongArrow(19,27)(29,37)
\Text(11,26)[b]{\small $k_1$}
\LongArrow(60,33)(70,43)
\Text(45,24)[tl]{\small $k_3$}
\LongArrow(60,-17)(70,-27)
\Text(45,-4)[tl]{\small $k_4$}

\end{picture}
\caption{\label{fig:BVR}
The lowest order amplitude (a), the IR divergent final-state virtual quark-pair correction (b) and the IR divergent real component (c).
The blob stands for the emission of $n\!-\!1$ particles.
Additional IR finite corrections are created if the gluon which splits into $q \bar q$ is emitted by the blob.}
\end{figure}
Our aim is to compute the large $N_F$ limit of cross sections including NNLO
quark-pair corrections. The relevant contributions are the Born, Virtual and Real reactions given by
\bqa
\label{eq:sigmaBVR}
\sigma_B &\propto& \int d \Phi_n\, \sum_{\rm spin}
|A^{(0)}_{n}|^2, \nl
\sigma_V &\propto& \int d \Phi_n\,
\sum_{\rm spin}
\left\{
 A^{(2)}_{n} (A^{(0)}_{n})^\ast 
+A^{(0)}_{n} (A^{(2)}_{n})^\ast 
\right\}, \nl
\sigma_R &\propto& \int d \Phi_{n+2}\,
\sum_{\rm spin}
\left\{
A^{(2)}_{n+2} (A^{(2)}_{n+2})^\ast
\right\}.
\eqa
In \eqref{eq:sigmaBVR}, $A^{(j)}_n$ represents the amplitude for the emission of $n$ partons computed at the $j^{th}$ perturbative order, while $d \Phi_m$ is the
$m$-particle phase-space
\bqa
d \Phi_m :=  \delta\left(P-\sum_{i=1}^m p_i\right) \prod^m_{i=1}
d^4 p_i \delta_+(p^2_i),
~~~{\rm with}~~~\delta_+(p^2_i):= \delta(p^2_i)\Theta(p^0_i), 
\eqa
in which $P$ is the initial state momentum.
The amplitude $A^{(0)}_{n}$ is drawn in figure~\ref{fig:BVR}-(a), where the line with momentum $p$ is an on-shell QCD parton and
the blob denotes $n$-1 additional final-state particles.
The NNLO amplitudes can be split into IR divergent and finite parts
\bqa
\label{eq:amplitudes}
A^{(2)}_{n} :=&& A^{(2)}_{n,\rm IR} + A^{(2)}_{n,\rm F},\nl
A^{(0)}_{n+2}:=&& A^{(0)}_{n+2,\rm IR} + A^{(0)}_{n+2,\rm F}.
\eqa
The infrared singular contributions are depicted in figure \ref{fig:BVR}-(b,c), while the finite pieces are created when the splitting gluon is emitted by the blob.

Due to the presence of IR parts in \eqref{eq:amplitudes},
both $\sigma_{V}$ and $\sigma_{R}$ are IR divergent, although, as is well known, their combination in infrared safe quantities is IR finite.
In addition, UV infinities are present in $\sigma_{V}$ that are renormalized away when bare parameters are determined at the perturbative order appropriate to match the NNLO accuracy.
In conclusion, the fully inclusive sum
\bqa
\sigma^{\mbox{\tiny\it NNLO}}= \sigma_{B}+\sigma_{V}+\sigma_{R}
\eqa
is a physical quantity, although its parts are separately plagued by IR and UV divergences.
Indeed, it is the simplest case of an infrared safe observable, which must give the same result when computed in any consistent scheme used to deal with the divergences.

In this section we use the four-dimensional FDR framework, and describe the procedures which allow one to compute $\sigma^{\mbox{\tiny\it NNLO}}$. We put particular emphasis on the steps needed to cope with the simultaneous presence of IR and UV infinities at two loops, and on how to merge virtual and real components.
Section \ref{sec:virt} presents the steps needed to define $\sigma_V$, while 
\ref{sec:real} deals with $\sigma_R$. Section \ref{sec:example}
contains an explicit example which guides the reader across both real and virtual procedures.

\subsection{The NNLO definition of the virtual contribution}
\label{sec:virt}
\begin{figure}[tbp]
\begin{picture}(215,110)(-220,-47)
\SetScale{1.5}
\SetOffset(-20,10)
\Line(-4,-4)(27,27)
\BCirc(-14,-14){17}
\SetScale{1}
\ArrowArc(37,-15)(15,-90,90)
\ArrowArc(37,-15)(15,90,270)
%
\GlueArc(-20,-14)(55,14,25){3}{1}
\GlueArc(-20,-14)(55,29,40){3}{1}
%
\GlueArc(-10,0)(43,-89,-62){3}{2}
\GlueArc(-10,0)(43,-55,-34){3}{2}
\DashLine(13,-27)(13,-47){2}
\Text(12,-50)[r]{$\hat \sigma$}
\DashLine(19,11)(40,11){2}
\Text(33,14)[lb]{$\hat \rho$}
\end{picture}
\caption{\label{fig:separated}
  The indices $\hat \rho$ and $\hat \sigma$ are external to the divergent
  sub-diagram disconnected form the rest.}
\end{figure}
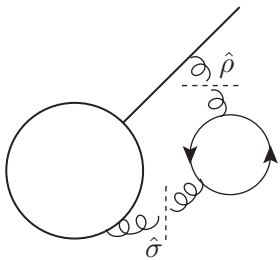

A generic two-loop integrand in $A^{(2)}_{n,\rm IR}$ has the form
\bqa
\label{eq:integrand}
J(q_1,q_2) =
\frac{F_{\hat \rho \hat \sigma}(q_1,q^2_1)}{D} \,\frac{
  q_2^{\rho} \qot^{\sigma} +\qot^\rho q_2^\sigma  -g^{\rho \sigma}
 (q_2 \cdot q_{12})}{q^2_2 \qots} := \frac{N}{D q^2_2 \qots},
\eqa
with internal loop momenta $q_1$, $q_2$ and $\qot := q_1 + q_2$.
The denominator $D$ collects all $q_1$-dependent propagators
\bqa
D= D_p q^4_1 \left(\varPi_{i=1}^k D_i\right),~~~~~~D_p = (q_1+p)^2,
\eqa
where $k$ is the number of propagators in the blob of figure~\ref{fig:BVR}-(b), and $N$ is the numerator of the integrand.
The hats on Lorentz indices means that they are external to the divergent sub-diagram, as in figure~\ref{fig:separated}. The difference between hatted and un-hatted indices can be ignored until equation \eqref{eq:FDRJ}.

We now analyze all possible divergences generated upon loop integration. We do this to determine the form of the FDR UV subtractions and the structure of the IR regulator
needed to define the FDR integration over \eqref{eq:integrand}  given in \eqref{eq:tildeI}.
$J(q_1,q_2)$ is quadratically divergent when $q_2 \to \infty$ at fixed $q_1$.
Given the presence of many propagators in $D$, this is the only possible UV sub-divergence. Depending on where the lower gluon reconnects to the blob, it may also generate global UV infinities. In addition, due to the on-shell condition $p^2= 0$, a double collinear IR singularity arises when both $q_1$ and $q_2$ are proportional to $p$. A second potential IR divergent configuration is $q^2_1 \to 0$ but $q^2_2 \ne 0$. Nevertheless, this divergence is cancelled by the UV behavior of the $q_2$ integration. This can be understood by noticing that a scaleless $q_2$-type sub-integral is generated in this case, that vanishes in FDR.~\footnote{See appendix \ref{app:B}.}
An additional double collinear IR singularity is created if the gluon is attached to a second external massless parton.

In FDR, an unphysical scale $\mu^2$ is added to all propagators in
order to regulate divergences. As is well known, performing this
operation only in the denominator is contrary to the core principle
\eqref{eq:prin:1} of gauge invariance. As such, in FDR one performs a {\em
Global Prescription} (GP) \cite{Pittau:2012zd}, also making the replacement in the
numerator such that all integrand cancellations between numerator and
denominator take place that the regulated level\footnote{This, combined with the shift invariance of FDR integrals is sufficient to prove all Ward identities graphically \cite{Donati:2013voa}.}. These cancellations are called ``gauge cancellations.''
In practice, one first determines the dependence of $J(q_1,q_2)$ on $q^2_1$, $q^2_2$ and $\qots$ generated by Feynman rules. This is the reason for the explicit $q^2_1$ as an argument of $F$ in \eqref{eq:integrand}.
Subsequently, one adds an unphysical scale $\mu^2$ to all such self-contractions
\bqa
\label{eq:gpre}
 q^2_i \to q^2_i -\mu^2 := \bar q^2_i. 
\eqa
We denote this procedure by the symbol $\toGP$. Thus, the replacement
$
A \toGP \bar A,
$
applied to any function $A(q^2_1, q^2_2, \qots)$, produces a new function
$\bar A$ with arguments replaced as in \eqref{eq:gpre}, 
\bqa
\bar A(\bar q^2_1, \bar q^2_2, \qotsb) := A(\bar q^2_1, \bar q^2_2, \qotsb).
\eqa
In the case at hand one has
\bqa
\label{eq:Jbar}
J(q_1,q_2) \toGP \bar  J(q_1,q_2) :=
\frac{\bar F_{\hat \rho \hat \sigma}(q_1,\bar q^2_1)}{\bar D} \bar G^{\rho \sigma}=
\frac{\bar N}{\bar D \bar q^2_2 \qotsb},
\eqa
with
\bqa
\bar D := \bar D_p \bar q^4_1 \left(\varPi_{i=1}^k \bar D_i\right),~~~
\bar D_{p,i} := D_{p,i}-\mu^2,
\eqa
and
\bqa
\label{eq:G1}
\bar G^{\rho \sigma} := \frac{
  q_2^{\rho} \qot^{\sigma} +\qot^\rho q_2^\sigma 
 -g^{\rho \sigma}
 (\qotsb + \bar q^2_2-\bar q^2_1)/2}{\bar q^2_2 \qotsb}.
\eqa

The IR singularities of $J(q_1,q_2)$ are now regulated by the addition of $\mu^2$ in the propagators. As we shall see, the asymptotic limit $\mu^2 \to 0$ will be eventually taken after integration. That generates
logarithms of $\mu^2$ of IR origin. Nevertheless $\bar  J(q_1,q_2)$ is still UV divergent. The global UV infinities are subtracted by separating physical and non-physical scales in $\bar D_p$. That means using the identity
\bqa
\label{eq:ident}
\frac{1}{\bar D_p}= \frac{1}{\bar q_1^2}
-\frac{2 (q_1 \cdot p)}{\bar q_1^2 \bar D_p}, 
\eqa
and noticing that the second term is more UV convergent than the original propagator. The same expansion has to be applied to the other propagators in $1/\bar D$, until $\bar F/\bar D$ is written as follows
\bqa
\label{eq:splitgv}
\frac{\bar F_{\hat \rho \hat \sigma}(q_1,\bar q^2_1)}{\bar D}=
 \left[\frac{\bar F_{\hat \rho \hat \sigma}(q_1,\bar q^2_1)}{\bar D}\right]_{V}
+\left(\frac{\bar F_{\hat \rho \hat \sigma}(q_1,\bar q^2_1)}{\bar D}\right)_{F},
\eqa
where $[\bar F/\bar D]_{V}$ do not depend on physical scales. Since also $\bar G^{\rho \sigma}$ does not contain physical scales, $[\bar F/\bar D]_{V}$ defines the global UV divergent behavior of $\bar  J(q_1,q_2)$:
\bqa
[\bar  J(q_1,q_2)]_{GV} := \left[\frac{\bar F_{\hat \rho \hat \sigma}(q_1,\bar q^2_1)}{\bar D}\right]_{V}\bar G^{\rho \sigma}.
\eqa
$[\bar  J(q_1,q_2)]_{GV}$ is called a {\em Global Vacuum} (GV) and is written between square brackets, that is the standard FDR notation to indicate the vacuum part of an object. Note that $(\bar F/\bar D)_{F}$ in \eqref{eq:splitgv} gives rise to a subtracted integrand which is globally UV convergent but still divergent when $q_2 \to \infty$:
\bqa
\left(\frac{\bar F_{\hat \rho \hat \sigma}(q_1,\bar q^2_1)}{ \bar D}\right)_{F} 
\bar G^{\rho \sigma}.
\eqa
This is fixed by subtracting the {\em Sub-Vacuum} (SV) from
$\bar G^{\rho \sigma}$ by means of the expansion 
\bqa
\label{eq:subsubt}
\frac{1}{\qotsb}=
\frac{1}{\bar q^2_2}-\frac{q^2_1+2 (q_1 \cdot q_2)}{\bar q^2_2 \qotsb}.
\eqa
The final result has the form
\bqa
\label{eq:vsub}
\bar G^{\rho \sigma}= \left[\bar G^{\rho \sigma}\right]_{SV}
            + \left(\bar G^{\rho \sigma}\right)_F,
\eqa
so that the fully UV subtracted integrand 
\bqa
\label{eq:intsub}
\left(
 \frac{\bar F_{\hat \rho \hat \sigma}(q_1,\bar q^2_1)}{\bar D}-
 \left[\frac{\bar F_{\hat \rho \hat \sigma}(q_1,\bar q^2_1)}{\bar D}\right]_{V}
 \right)
 \left(
 \bar G^{\rho \sigma}-\left[\bar G^{\rho \sigma}\right]_{SV}
\right),
\eqa
is integrable in four dimensions. Upon integration, the vacua subtracted in \eqref{eq:intsub} induce the appearance of logarithms of $\mu^2$ of UV origin, so that both IR and UV singularities are regulated by the same regulator.

The procedure leading to \eqref{eq:intsub} is conveniently encoded in a linear integral operator
\bqa
\label{eq:FDRope}
\int [d^4q_1] [d^4q_2],  
\eqa
whose action on a two-loop integrand is defined by three subsequent operations:
\begin{itemize}
  \item subtract the vacua;
  \item integrate over $q_1$ and $q_2$; 
  \item take the asymptotic limit $\mu^2 \to 0$. 
\end{itemize}
The last operation means retaining only the logarithmic pieces in the asymptotic expansion, neglecting ${\cal O}(\mu^2)$ terms.
Thus, the FDR two-loop integration over $\bar J(q_1,q_2)$ in \eqref{eq:Jbar} is defined as follows
\bqa
\label{eq:FDRJ}
\bar I := \int [d^4q_1] [d^4q_2]
%
\frac{\bar N}{\bar D \bar q^2_2 \qotsb}
=
\int d^4q_1 d^4q_2
\left(\frac{\bar F_{\hat \rho \hat \sigma}(q_1,\bar q^2_1)}{ \bar D}
\right)_F
\left(\bar G^{\rho \sigma}
\right)_F.
\eqa
In the following, we will often omit terms that integrate to zero in globally prescribed numerators. Hence, it is convenient to introduce a notation for that, that is
$
\bar N^\prime  \simeq \bar N
$
if both numerators give the same result upon FDR integration.

\begin{figure}[tbp]
  \scalebox{0.8}{
  \begin{picture}(215,110)(-100,-47)
    \SetScale{1.5}
    \SetOffset(-110,10)
    \Line(-4,-4)(27,27)
    \BCirc(-14,-14){17}
    \SetScale{1}
    \DashLine(17,11)(65,11){2}
    \Text(10,11)[l]{$\lhd$}
    \Text(65,11)[l]{$\otimes$}
    \Text(90,-21)[]{\LARGE$+$}
    \SetScale{1.5}
    \SetOffset(50,10)
    \Line(-4,-4)(27,27)
    \BCirc(-14,-14){17}
    \SetScale{1}
    \DashLine(10,-21)(55,-21){2}
    \Text(3,-21)[l]{$\lhd$}
    \Text(55,-21)[l]{$\otimes$}
    \Text(90,-21)[]{\LARGE$= 0$}
  \end{picture}
  }
  \centering
  \caption{\label{fig:graphycalWI}
    The graphical version of the WI in \eqref{eq:2wi}. The scalar gluon, with coupling proportional to the gluon momentum, is denoted by a dashed arrow.
    The symbol $\otimes$ indicates that it is emitted by the quark loop.}
  \end{figure}
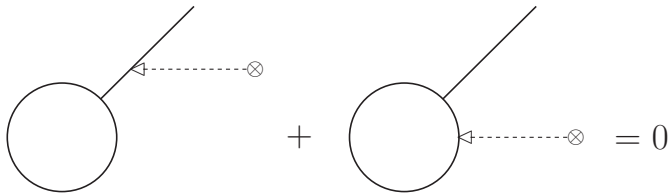

Equation \eqref{eq:FDRJ} defines a gauge-invariant object, in which the necessary gauge cancellations are preserved by the GP operation. By ``gauge-invariant object'' we mean that a calculation of $\bar{I}$ in a different gauge will give the same result. It is instructive to check that no change in $\bar{I}$ is produced if one shifts the numerator of the gluon propagator as
\bqa
g_{\rho \hat \rho} \to   g_{\rho \hat \rho} + \lambda_1 \frac{q^\rho_1 q^{\hat \rho}_1}{\bar q^2_1},~~~~
g_{\sigma \hat \sigma} \to   g_{\sigma \hat \sigma} + \lambda_2 \frac{q^\sigma_1 q^{\hat \sigma}_1}{\bar q^2_1},
~~~~\forall \lambda_1, \lambda_2.
\eqa
Thus, $\bar I$ gives the same result when computed in any gauge. Another consequence of the WIs
is that the term proportional $q_2^\rho q_1^\sigma+q_1^\rho q_2^\sigma$
in \eqref{eq:G1} should not contribute to  $\bar I$ when contracted with $\bar F_{\hat \rho \hat \sigma}$. That is, 
\bqa
\label{eq:2wi}
\int [d^4q_1] [d^4q_2]
\frac{\bar F_{\hat \rho \hat \sigma}(q_1,\bar q^2_1)}{\bar D \bar q^2_2 \qotsb} \left(q_1^\rho q_2^\sigma + q_2^\rho q_1^\sigma \right)=0.
\eqa
After taking into account the vanishing of scaleless integrals, this corresponds to the WI depicted in figure \ref{fig:graphycalWI}.
Nevertheless, we keep this piece in \eqref{eq:G1}, as we will explicitly show in our calculation that it never contributes.

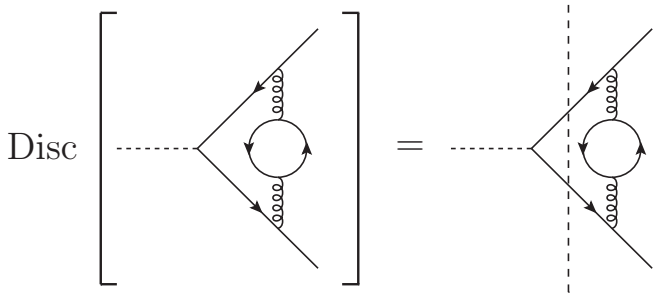
\begin{figure}[t]
  \begin{picture}(215,110)(-215,-80)
    \SetScale{0.6}
    \SetWidth{1}
    \SetOffset(-30,-30)
    \Text(-75,1)[r]{\Large Disc}
    \ArrowArc(0,0)(18,-90,90)
    \ArrowArc(0,0)(18,90,270)
    \Gluon(0,-18)(0,-50){3}{4}
    \Gluon(0, 50)(0, 18){3}{4}
    \DashLine(-50,0)(-100,0){3}
    \ArrowLine(25,75)(-50,0)
    \ArrowLine(-50,0)(25,-75)
    \SetWidth{1.5}
    \Line(-110,-85)(-110,85)
    \Line(50,-85)(50,85)
    \Line(-110,-85)(-100,-85)
    \Line(-110,85)(-100,85)
    \Line(50,-85)(40,-85)
    \Line(50,85)(40,85)
    \SetWidth{1}
    \SetOffset(95,-30)
    \Text(-70,0)[r]{\Large =}
    \ArrowArc(0,0)(18,-90,90)
    \ArrowArc(0,0)(18,90,270)
    \Gluon(0,-18)(0,-50){3}{4}
    \Gluon(0, 50)(0, 18){3}{4}
    \DashLine(-50,0)(-100,0){3}
    \ArrowLine(25,75)(-50,0)
    \ArrowLine(-50,0)(25,-75)
    \DashLine(-27,-90)(-27,90){5}
  \end{picture}
  \caption{\label{fig:uniSIC}
    Perturbative expansion of \eqref{eq:unitarity} for $H \to b \bar b + jets$
    in the large $N_F$ limit.}
  \end{figure}

Let us now consider the unitarity properties of this prescription.
Equation \eqref{eq:unitarity} relates different orders of
perturbation theory, so a given divergent subgraph $D$ can appear
embedded in loop diagrams of different orders $\ell$. If the result of
$D$ depends on $\ell$ then the relation \eqref{eq:unitarity} will in
general not survive. For a more concrete example relevant to the
calculations of this paper, consider the consequence of
\eqref{eq:unitarity} shown in figure \ref{fig:uniSIC}. This specifies
that the discontinuity of a two-loop graph is given in terms of a
one-loop graph. As such, the higher loop regularization must be
consistent with the lower loop one. In FDR this places strong
constraints on the GP. In our example of figure \ref{fig:uniSIC} we
see that on the left hand side, the momenta associated to the gluon
lines take part in the GP, but this is not true for the right hand
side. In FDR, this tension is resolved by enforcing
``sub-integration consistency''(SIC)
\cite{Page:2015zca}. 

We now consider this procedure in detail for the integrand
$J(q_1, q_2)$.  We analyze a subset of the full integrand,
specifically the numerator of the gluonic self-energy of figure
\ref{fig:separated},
$
q_2^{\rho} \qot^{\sigma} +\qot^\rho q_2^\sigma 
 -g^{\rho \sigma}
 (q_2 \cdot \qot)
$, 
which appears in \eqref{eq:integrand}.
This is the piece that needs to be treated carefully to maintain SIC. We consider, in particular, the result of the contraction with a
$g_{\hat \rho \hat \sigma}$\footnote{The analysis for terms like $\gamma_{\hat \rho}\gamma_{\hat \sigma}$ is also required, but equivalent.} (potentially) contained in $F_{\hat \rho \hat \sigma}$. It reads
\bqa
N_s(q^2_1,q^2_2,\qots, \hat q^2_2) :=  q^2_1 -3q^2_2 -\qots + 2 \hat q^2_2,
\eqa
where we have explicitly separated the contribution
$
\hat q^2_2 :=g_{\hat \rho \hat \sigma} q^\rho_2 q^\sigma_2
$.
From the point of view of the sub-diagram disconnected from the rest, $\hat q^2_2$ should not be globally prescribed, hence the GP replacement to be performed is
\bqa
\label{eq:inc0}
N_s(q^2_1,q^2_2,\qots, \hat q^2_2) \to N_s(q^2_1,\bar q^2_2,\qotsb, q^2_2). 
\eqa
On the other hand, embedding this in the full diagram requires integrating over $q_1$, therefore $q^2_1$ needs to be barred
\bqa
\label{eq:inc1}
N_s(q^2_1,q^2_2,\qots, \hat q^2_2) \to N_s(\bar q^2_1,\bar q^2_2,\qotsb, q^2_2).
\eqa
Nevertheless, $\rho$ and $\sigma$ are internal indices of the two-loop diagrams of figure \ref{fig:BVR}-(b), so that GP is needed for $\bar q^2_2$ as well
\bqa
\label{eq:inc2}
N_s(q^2_1,q^2_2,\qots, \hat q^2_2) \toGP N_s(\bar q^2_1,\bar q^2_2,\qotsb, \bar q^2_2),
\eqa
which is the prescription used in \eqref{eq:FDRJ}.
It is the mismatch among \eqref{eq:inc0}, \eqref{eq:inc1} and \eqref{eq:inc2} which violates SIC. Our solution to this problem is modifying the integrand in 
\eqref{eq:FDRJ} as follows:
\begin{itemize}
\item we do not apply GP to $\hat q^2_2$ terms whose origin is a contraction with indices external to the UV divergent sub-diagram;
\item we replace back everywhere $\bar q^2_1 \to q^2_1$ {\em after} GV subtraction.
\end{itemize}
Note that the last operation is possible because barring $q^2_2$ and
$\qots$ is sufficient to regulate the IR divergences. This is a
consequence of the fact that the only IR divergent configuration is
the double collinear limit. Furthermore, this solution does not affect
any of the WIs and so still defines a gauge invariant
object.
In summary, after GV subtraction, \eqref{eq:inc0} is maintained as it is also when embedded
in a two-loop calculation.
A comparison between this solution and the IR-free case in given in appendix~\ref{app:A}. 

Let us now return to the matter of enforcing SIC in the entirety of \eqref{eq:FDRJ}, and how one applies our solution. We enforce SIC by rewriting
\eqref{eq:FDRJ} as
\bqa
\bar I= 
\int [d^4q_1] [d^4q_2]
\left(\frac{\bar F_{\hat \rho \hat \sigma}(q_1,\bar q^2_1)}{\bar D}\right)_F
\bar G^{\rho \sigma}(\bar q^2_1),
\eqa
where we have used the fact that $[\bar F/\bar D]_V$ is subtracted by the integral operator.
The structure of the expansions needed to extract the GV is such that $\bar D$ can be always pulled out from the rest. Thus, it is possible to rewrite
\bqa
\left(\frac {\bar F_{\hat \rho \hat \sigma}(q_1,\bar q^2_1)}{\bar D}\right)_F = \frac{\bar H_{\hat \rho \hat \sigma}(q_1,\bar q^2_1)}{\bar D}.
\eqa
Next, we introduce the numerator function
\bqa
\label{eq:barN}
{\bar {\cal Z }}(\bar q^2_1,\hat q^2_2) := \bar H_{\hat \rho \hat \sigma}(q_1,\bar q^2_1) \bar G^{\rho \sigma}(\bar q^2_1)
\bar q^2_2 \qotsb,
\eqa
where the explicit dependence on $\hat q^2_2$ is generated by the l.h.s of \eqref{eq:inc0}.
In practice, ${\bar {\cal  Z}}(\bar q^2_1,\hat q^2_2)$ is constructed from $N$ in \eqref{eq:integrand} as follows:
\begin{itemize}
\item globally prescribe $N$,
  $ N \toGP \bar N$,
  leaving $\hat q^2_2$ unbarred;
\item perform GV subtraction and determine, for each term $T$ in $\bar N$, the appropriate function $\bar H$ to be used in \eqref{eq:barN}. The result of this will always have a factorized form. For instance, if \eqref{eq:ident} has to be used once to subtract the global vacuum, the  contribution of $T$ to
  ${\bar {\cal  Z}}(\bar q^2_1,\hat q^2_2)$ is
  $-2(q_1 \cdot p)/\bar q^2_1 \times T$;
\item eliminates the bars from the $\bar q^2_1$s;
\item identify $\hat q^2_2$ with $q^2_2$.
\end{itemize}
We denote the last two operations with the symbol $\toSIC$.
Thus, the change
$
\bar A \toSIC \tilde A,
$
applied to any globally prescribed and GV subtracted function
$\bar A(\bar q^2_1,\hat q^2_2)$, produces a new function defined as
\bqa
\tilde A(q^2_1,q^2_2) := \bar A(q^2_1,q^2_2).
\eqa
Thus, the SIC compatible version of \eqref{eq:barN} reads
\bqa
\label{eq:barT}
\bar {\cal Z}(\bar q^2_1,\hat q^2_2) \toSIC \tilde {\cal Z}(q^2_1,q^2_2).
\eqa
To continue to ensure gauge cancellations, 
we unbar also the propagators, that leads to
\bqa
\label{eq:tildeI}
\tilde I := 
\int d^4q_1 [d^4q_2] \frac{\tilde {\cal Z}(q^2_1,q^2_2)}{D \bar q^2_2 \qotsb}.
\eqa
Equation \eqref{eq:tildeI} defines the SIC preserving four-dimensional integration over the integrand in \eqref{eq:integrand}. Note that, since the GV has been subtracted in it, $[d^4q_1]$ is replaced by a customary integration $d^4q_1$.
The asymptotic limit $\mu^2 \to 0$ is understood after taking the two integrations.

A first consequence of this definition is that external wave-function corrections vanish for massless particles, so that they can always be neglected in actual calculations. The proof is given in appendix~\ref{app:B}.

\subsection{The NNLO definition of the real component}
\label{sec:real}
Given the propagator structure of figure~\ref{fig:BVR}-(c), the integrands
contributing to $\sigma_R$ in \eqref{eq:sigmaBVR} have the following form
\bqa
\label{eq:Jr}
J_R = \frac{N_R}{S s^\alpha_{34} s^\beta_{134}},~~~s_{i\ldots j} := (k_i+\ldots + k_j)^2,~~~0 \le \alpha,\beta\le 2, 
\eqa
where $S$ collects all the remaining propagators and $N_R$ is the numerator of the amplitude squared.
Depending on the value of the exponents $\alpha$ and $\beta$, 
$J_R$ becomes IR divergent under integration over $\Phi_{n+2}$. These IR singularities must be regulated coherently with our treatment of the virtual component, without violating unitarity and gauge invariance. In this section, we determine a four-dimensional integration that achieves this.

Our starting point is the representation of $\sigma_V$ and $\sigma_R$ in terms of cut diagrams, in which we put the complex conjugate amplitudes on the right side. With this convention,
normal Feynman rules are assumed on the left and complex conjugate ones on the right. 
\begin{figure}[tbp]
\begin{picture}(215,135)(-160,-65)
\SetScale{1}
\SetOffset(-110,0)
\Line(-35,40)(35,40)
\Gluon(0,15)(0,40){3}{3}
\ArrowArc(0,0)(15,-90,90)
\ArrowArc(0,0)(15,90,270)
\Gluon(0,-40)(0,-15){3}{3}
\DashLine(22,55)(22,-40){3}
\LongArrow(10,45)(30,45)
\Text(14,50)[b]{\small $p$}
\LongArrow(-33,45)(-13,45)
\Text(-22,50)[b]{\small $q_1\!+\!p$}
\Text(0,-55)[t]{(a)}
\Text(10,0)[r]{\small $q_2$}
\Text(-20,0)[r]{\small $\qot$}
\SetOffset(0,0)
\Line(-35,40)(35,40)
\Gluon(0,15)(0,40){3}{3}
\ArrowArc(0,0)(15,-90,90)
\ArrowArc(0,0)(15,90,270)
\Gluon(0,-40)(0,-15){3}{3}
\DashLine(-22,55)(-22,-40){3}
\Text(0,-55)[t]{(b)}
\SetOffset(110,0)
\Line(-35,40)(35,40)
\Gluon(0,15)(0,40){3}{3}
\ArrowArc(0,0)(15,-90,90)
\ArrowArc(0,0)(15,90,270)
\Gluon(0,-40)(0,-15){3}{3}
\DashLine(37,55)(-27,-40){3}
\LongArrow(20,45)(40,45)
\Text(25,50)[b]{\small $k_1$}
\Text(0, 43)[b]{\small $\hat \rho$}
\Text(0,-43)[t]{\small $\hat \sigma$}
\LongArrow(7,18)(17,11)
\LongArrow(-17,-11)(-7,-18)
\Text(22,11)[l]{\small $k_3$}
\Text(-20,-11)[r]{\small $k_4$}
\Text(0,-55)[t]{(c)}
\SetOffset(220,0)
\Line(-35,40)(35,40)
\Gluon(0,15)(0,40){3}{3}
\ArrowArc(0,0)(15,-90,90)
\ArrowArc(0,0)(15,90,270)
\Gluon(0,-40)(0,-15){3}{3}
\DashLine(-37,55)(27,-40){3}
\Text(0, 43)[b]{\small $\hat \rho$}
\Text(0,-43)[t]{\small $\hat \sigma$}
\LongArrow(-40,45)(-20,45)
\Text(-24,50)[b]{\small $k_1$}
\LongArrow(-17,11)(-7,18)
\LongArrow(7,-18)(17,-11)
\Text(-19,11)[r]{\small $k_3$}
\Text(22,-11)[l]{\small $k_4$}
\Text(0,-55)[t]{(d)}
\Text(0,0)[]{\huge .}
\end{picture}
\caption{\label{fig:cuts} Virtual and real cuts contributing to the IR divergent parts of $\sigma_V$ (a,b) and $\sigma_R$ (c,d).}
\end{figure}
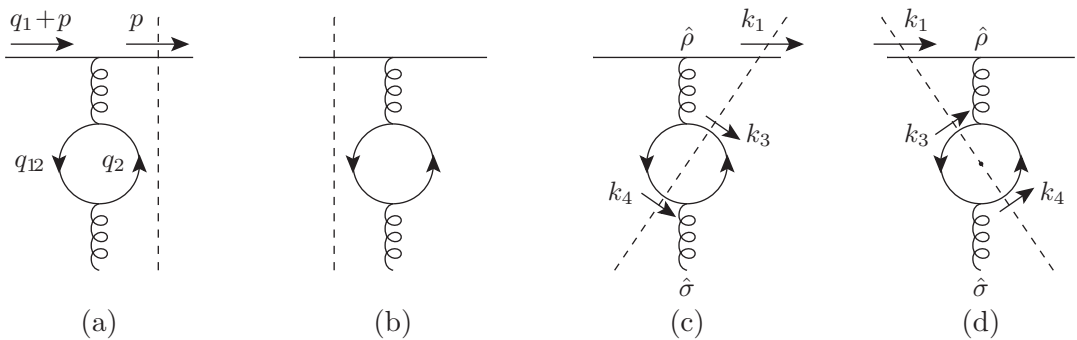
The cuts generating IR divergent configurations are obtained by squaring the amplitudes in figure \ref{fig:BVR}-(b,c) and are depicted in figure~\ref{fig:cuts}. IR divergences manifest themselves as pinch singularities of the loop integrals in (a,b) and endpoint phase-space singularities in (c,d). 
These two kinds of singularities are  related to each other by the identity
\bqa
\label{eq:propid}
\frac{1}{k^2+ i0^+}= \frac{2 \pi}{i} \delta_+(k^2)+\frac{1}{k^2+i k^0 0^+},
\eqa
in which the poles of the propagator on the l.h.s. may create a pinch in the complex plane of the loop integration and the $\delta_+(k^2)$ on the  r.h.s. may induce a non-integrable end-point configuration.  
Dubbing $\Sigma_{c}$ the sum over all cuts that appears in the r.h.s. of \eqref{eq:unitarity}, the cutting equations
\cite{Cutkosky:1960sp,Veltman:1963th} ensure that the last term in \eqref{eq:propid} does not contribute
to the singular part of each cut in $\Sigma_{c}$, and that $\Sigma_{c}$ is IR finite.
This theorem implies a unitarity-preserving cancellation of the IR singularities if the Cutkosky relation
\begin{subequations}\label{eq:cond}
\bqa
\label{eq:cond:1}
\frac{1}{k^2+ i0^+} \leftrightarrow \frac{2 \pi}{i} \delta_+(k^2),
\eqa
giving the possibility of a one-to-one integrand level identification of the infrared divergent parts contributing to different cuts in $\Sigma_{c}$, is preserved. However, one should also prove that $\Sigma_{c}= \sigma_R + \sigma_V$.
The reason for this second requirement is the different origin of the potential numerators multiplying the two sides of \eqref{eq:cond:1}.
In the case of a fermion line, that is the only cut relevant for this paper, the
l.h.s. gets multiplied by the numerator of the propagator
$\rlap/k:= \rlap/k_{\rm prop}$, while the r.h.s. by  
$\sum_{\rm spin}u(k)\bar u(k)=\sum_{\rm spin}v(k)\bar v(k):= \rlap/k_{\rm spin}$.
Hence, in addition to \eqref{eq:cond:1}, the identity
\bqa
\label{eq:cond:2}
\rlap/k_{\rm prop} = \rlap/ k_{\rm spin},
\eqa
\end{subequations}
must hold to guarantee the validity of \eqref{eq:unitarity}. Note that \eqref{eq:cond:2} also guarantees consistent gauge cancellations in all terms contributing to $\Sigma_{c}$.

Let us consider how to make these relations consistent with the procedure from the previous section where we make two modifications to the integrand:
\begin{itemize}
\item adding $\mu^2$ to a few propagators;
\item SV and GV subtraction from the integrand.
\end{itemize}

We first study how FDR preserves \eqref{eq:cond:1}. We start dealing with the effect of the  $q_2^2 \toGP \bar q^2_2$ and $\qots \toGP \qotsb$ replacements in cuts (a,b). Equation \eqref{eq:cond:1} is preserved if
\begin{equation}
  \label{eq:qtok}
\frac{1}{(\bar q_2^2+i0^+)(\qotsb+i0^+)} \leftrightarrow
\left(\frac{2 \pi}{i}\right)^2
\delta_+(\bar k^2_3)
\delta_+(\bar k^2_4),~~~{\rm with} ~~~\bar k^2_{3,4}:= k^2_{3,4}-\mu^2.
\end{equation}
Thus, the $\bar q_2^2$ and $\qotsb$ propagators in $\sigma_V$ must correspond to external particles in $\sigma_R$ obeying
\bqa
\label{eq:k34b}
k^2_{3,4}=\mu^2.
\eqa
Hence, we replace in \eqref{eq:sigmaBVR}
$\Phi_{n+2} \to \tilde \Phi_{n+2}$, where the phase-space $\tilde \Phi_{n+2}$ is such that $k^2_3=k^2_4= \mu^2$ and $k^2_i= 0$ when $i \ne 3,4$.
However, this is not enough. One also needs to show that \eqref{eq:qtok} survives the SV subtraction of \eqref{eq:intsub}.
We prove this explicitly in the case of the last term in \eqref{eq:G1}. The proof is unchanged for the other contributions. The relevant expansion is
\bqa
\label{eq:svss}
\frac{1}{\bar q^2_2 \qotsb}=
\left[\frac{1}{\bar q^2_4}\right]_{SV}-\frac{q^2_1+2 (q_1 \cdot q_2)}{\bar q^4_2 \qotsb},
\eqa
where SV is the term to be subtracted.
We consider a piece of the finite part of \eqref{eq:svss}
as a numerator factor $f$  
\bqa
-\frac{q^2_1+2 (q_1 \cdot q_2)}{\bar q^4_2 \qotsb}= \frac{f}{\bar q^2_2 \qotsb},~~~f:= -\frac{q^2_1+2 (q_1 \cdot q_2)}{\bar q^2_2}
\eqa
and observe that $f \to 1$ when $\qotsb \to 0$.
We therefore first put the $\qotsb$ propagator on-shell, giving
\bqa
\frac{1}{\bar q^2_2 \qotsb}
\leftrightarrow \frac{(2 \pi)}{i} \delta_+(\bar k^2_4)
\left\{
 \frac{2(k_3 \cdot k_{34})-k^2_{34}}{\bar k^4_3}
  \right\}=
\frac{(2 \pi)}{i} \frac{\delta_+(\bar k^2_4)}{\bar k^2_3}.
\eqa
When also $\bar k^2_3$ goes on-shell, one obtains the same result 
as applying \eqref{eq:qtok} before subtracting the vacuum.
Hence, the SV subtraction is ``invisible'' from the point of view of
\eqref{eq:qtok}, and \eqref{eq:cond:1} is fulfilled if $k_{3,4}$ obey
\eqref{eq:k34b}.

 As for \eqref{eq:cond:2}, in order to preserve it, one must treat $\rlap /k_3$ and $\rlap /k_4$ in the numerator $N_R$ of \eqref{eq:Jr} using the same prescriptions imposed on 
$\rlap/q_2$ and $\rlap/ \qot$ in $N$. This means replacing in $N_R$
\bqa
\label{eq:gprer}
k^2_{3,4} \to \bar k^2_{3,4} = 0,~~~
(k_3 \cdot k_4) \to \frac{1}{2}s_{34}.
\eqa
These changes should be performed everywhere in $N_R$ except in contractions induced by the external indices $\hat \rho$ and $\hat \sigma$ in cuts (c,d). In this case 
\bqa
\label{eq:nogpr}
\hat k_{3,4}^2= k_{3,4}^2= \mu^2,~~~
(\hat k_3 \cdot \hat k_4) = (k_3 \cdot k_4) = \frac{s_{34}-2 \mu^2}{2},
\eqa
in accordance to the SIC preserving requirement we have used to construct ${\tilde {\cal  Z}}(q^2_1,q^2_2)$.
We denote all of this by introducing a globally prescribed and SIC preserving version of $N_R$
\bqa
\label{eq:NRtilde}
N_R \toGS \tilde N_R(\mu^2), 
\eqa
where the action of $\toGS$ on a function 
$
A(k^2_3,k^2_4,\hat k^2_3, \hat k^2_4) 
$
is defined to be
\bqa
\label{eq:GS}
A(k^2_3,k^2_4,\hat k^2_3, \hat k^2_4)
\toGS
  \tilde A(\bar k^2_3, \bar k^2_4, k^2_3, k^2_4) 
:=       A(\bar k^2_3, \bar k^2_4, k^2_3, k^2_4). 
\eqa
Only a dependence on $\mu^2$ is left in the r.h.s. of \eqref{eq:NRtilde}
because of the deltas in  $\tilde \Phi_{n+2}$.

The only remaining propagator modified by our definition of the virtual component is in 
the top-left line of cut (a), that must correspond to the cut $k_1$ propagator  in (d)
through the relation~\footnote{The complex conjugate of \eqref{eq:lastprop} links cuts (b) and (c).}
\bqa
\label{eq:lastprop}
\frac{1}{(q_1+p)^2+i0^+}  \leftrightarrow  \frac{2 \pi}{i} \delta_+(k^2_1).
\eqa
Everything is massless in this case, so that \eqref{eq:cond:2} is fulfilled and it is sufficient to  check that GV subtraction does not alter \eqref{eq:lastprop}. The proof is similar to the one used for the SV. In fact, if $m$ expansions
\bqa
\frac{1}{(q_1+p)^2-\mu^2}= \frac{1}{\bar q_1^2}
-\frac{2 (q_1 \cdot p)}{\bar q_1^2 ((q_1+p)^2-\mu^2)} 
\eqa
are needed to subtract the vacuum in front of a term in
$\bar {\cal Z}(\bar q^2_1,\hat q^2_2)$, this term gets multiplied by a factor $f^m$ in $\tilde {\cal Z}(q^2_1,q^2_2)$, where 
\bqa
f := \frac{-2(q_1 \cdot p)}{q^2_1}.
\eqa
But $f=1$ when the propagator goes on-shell. So that \eqref{eq:lastprop} survives GV subtraction.

In summary, we define the four-dimensional integration over the integrand in
\eqref{eq:Jr} as follows
\bqa
\label{eq:rint}
\tilde R := \int d \tilde \Phi_{n+2} \frac{\tilde N_R(\mu^2)}{S s^\alpha_{34} s^\beta_{134}}.
\eqa
By doing that, unitarity preserving IR cancellations occur by construction between $\sigma_R$ and $\sigma_V$, without violating gauge invariance.

\subsection{An example of cancellation}
\label{sec:example}
The integrals in \eqref{eq:tildeI} and \eqref{eq:rint}
can be computed independently, and this is the strategy we adopt in this paper.
In fact, term by term cancellations in $\Sigma_c$ are difficult to find. The reason is that one can add to the numerator of the virtual piece arbitrary vanishing terms that nevertheless contribute to the real part, and vice-versa. This is due to the different structure of the deltas contained in $d \Phi_{n}$ and
$d \tilde \Phi_{n+2}$. However, if the numerator of a term does not change - modulo a relabelling of the momenta - when multiplied by both phase-spaces, the IR cancellation must occur between integrals constructed one from the other via the replacement in \eqref{eq:cond:1}.
In this section we illustrate this phenomenon by considering a piece of the full $H \to b \bar b + {jets}$ calculation presented in the following section. This also serves us as a concrete example on how the procedures of sections
\ref{sec:virt} and \ref{sec:real} work in practice. 

\begin{figure}[tbp]
\begin{picture}(215,160)(-215,-80)
\SetScale{1}
\SetWidth{1}
\SetOffset(0,0)
\ArrowArc(0,0)(50,0,100)
\ArrowArc(0,0)(50,90,180)
\ArrowArc(0,0)(50,180,270)
\ArrowArc(0,0)(50,260,360)
\SetWidth{1}
\ArrowArc(0,0)(20,-90,90)
\ArrowArc(0,0)(20,90,270)
\SetWidth{1}
\Gluon(0,-20)(0,-50){3}{4}
\Gluon(0, 50)(0, 20){3}{4}
\DashLine(50,0)(100,0){3}
\DashLine(-50,0)(-100,0){3}
\DashLine(40,-70)(40,70){5}
\SetWidth{0.25}
\LongArrow(7,42)(7,28)
\Text(9,35)[l]{$q_1$}
\LongArrow(7,-28)(7,-42)
\Text(9,-35)[l]{$q_1$}
\Text(-3,58)[l]{$\hat \rho$}
\Text(-3,14)[l]{$\rho$}
\Text(-3,-14.5)[l]{$\sigma$}
\Text(-3,-58)[l]{$\hat \sigma$}

\LongArrow(25,-7)(25,7)
\Text(27,-4)[l]{$q_2$}
\LongArrow(-25,7)(-25,-7)
\Text(-27,2)[r]{$q_{1\!2}$}
\LongArrow(32,48)(48,32)
\Text(46,43)[l]{$p_2$}
\LongArrow(48,-32)(32,-48)
\Text(46,-43)[l]{$p_1$}
\LongArrow(-48,32)(-32,48)
\Text(-45,45)[r]{$q_1\!+\!p_2$}
\LongArrow(-32,-48)(-48,-32)
\Text(-45,-45)[r]{$q_1\!+\!p_1$}
\LongArrow(62,5)(72,5)
\Text(68,9)[b]{$P$}

\end{picture}
\caption{\label{fig:1}
A cut contributing to $H \to b \bar b$ at NNLO. Only the term
proportional to $g_{\rho \sigma}$ is considered in \eqref{eq:gammav}.}
\end{figure}
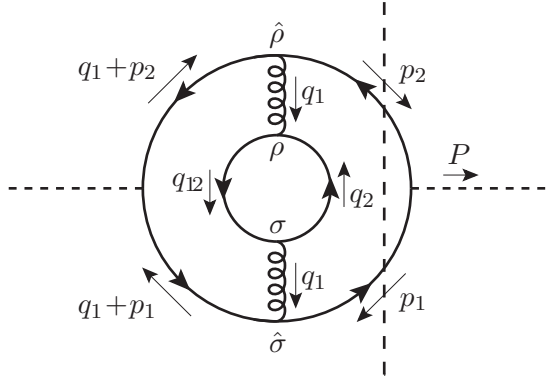

The two-loop contribution to $\sigma_V$ is depicted in figure~\ref{fig:1}.
We focus on the term generated by the $g_{\rho \sigma}$ piece of the internal trace.  It reads
\bqa
\label{eq:gammav}
\bar \Gamma_V = -\frac{1}{(2 \pi)^{10}}\int d^4{p_1} d^4{p_2} [d^4{q_1}] [d^4{q_2}]
\delta_-(p^2_1) \delta_+(p^2_2)
\delta^4(P-p_2+p_1)
\frac{\bar N_V}{\bar q_1^4  \bar D_1 \bar D_2 \bar q_2^2 \qotsb}, 
\eqa
with ${D_i:=(q_1+p_i)^2}$.
The numerator ${\bar N_V}$ is obtained via GP from its unbarred version
\bqa
\label{eq:NVunb}
N_V = -64 (p_1 \cdot p_2) (q_2 \cdot \qot )(q_1+p_1) \!\cdot\! (q_1+p_2),
\eqa
where we have neglected couplings and color factors, but not phases needed to compute the overall sign.
We rewrite
\bqa
N_V = 16 s (\qots+q^2_2-q^2_1)\left(D_1+(P \cdot q_1 )-\frac{s}{2} \right),
\eqa
with $s= P^2$, that gives
\bqa
\label{eq:NV}
N_{V} \toGP \bar N_{V} =16 s (\qotsb+\bar q^2_2-\bar q^2_1)\left(\bar D_1+(P \cdot q_1)-\frac{s}{2} \right).
\eqa
This expression can be simplified by dropping terms which integrate to zero, for example $\bar q^2_2$ and $\qotsb$ generate vacua and $\bar D_1$ gives a scale-less integral. Furthermore, $(P \cdot q_1 )$ does not contribute because it is antisymmetric when $p_1 \leftrightarrow p_2$, while the denominator in \eqref{eq:gammav} is symmetric. Thus $\bar N_{V} \simeq \bar N^\prime_{V}$, with
\bqa
\label{eq:NVfinal}
\bar N^\prime_{V} (\bar q^2_1) = 8 s^2 \bar q^2_1.
\eqa
The GV in $\eqref{eq:gammav}$ is fully removed by the subtraction of the scale-less integral. Thus,  ${\bar {\cal  Z}_V}(\bar q^2_1)= \bar N^\prime_{V}(\bar q^2_1 )$ and
\bqa
\label{eq:NVex}
{\bar {\cal  Z}_V}(\bar q^2_1) \toSIC {\tilde {\cal  Z}_V}(q^2_1) =  \bar N^\prime_{V} (q^2_1)= 8 s^2 q^2_1,
\eqa
so that the physically relevant two-loop contribution reads
\bqa
\label{eq:gammavt}
\tilde \Gamma_V &=& \frac{1}{(2 \pi)^{10}}\int d^4{p_1} d^4{p_2} d^4{q_1} [d^4{q_2}]
\delta_-(p^2_1) \delta_+(p^2_2)
\delta^4(P-p_2+p_1)
\frac{{\tilde {\cal  Z}_V}(q^2_1)}{q_1^4 D_1 D_2 \bar q_2^2 \qotsb}.
\eqa

\begin{figure}[tbp]
\begin{picture}(215,160)(-215,-80)
\SetScale{1}
\SetWidth{1}
\SetOffset(0,0)
\ArrowArc(0,0)(50,-40,90)
\ArrowArc(0,0)(50,90,180)
\ArrowArc(0,0)(50,140,270)
\ArrowArc(0,0)(50,270,360)
\SetWidth{1}
\ArrowArc(0,0)(20,-90,90)
\ArrowArc(0,0)(20,90,270)
\SetWidth{1}
\Gluon(0,-20)(0,-50){3}{4}
\Gluon(0, 50)(0, 20){3}{4}
\DashLine(50,0)(100,0){3}
\DashLine(-50,0)(-100,0){3}
\DashLine(70,70)(-70,-70){5}
\SetWidth{0.25}
\LongArrow(7,-28)(7,-42)
\Text(10,-35)[l]{$k_{34}$}
\LongArrow(10,26)(26,10)
\Text(16,28)[br]{$k_3$}
\LongArrow(-26,-10)(-10,-26)
\Text(-24,-8)[br]{$k_4$}
\Text(-3,58)[l]{$\hat \rho$}
\Text(-3,14)[l]{$\rho$}
\Text(-3,-14.5)[l]{$\sigma$}
\Text(-3,-58)[l]{$\hat \sigma$}
\LongArrow(32,48)(48,32)
\Text(34,50)[br]{$k_2$}
\LongArrow(32,-48)(48,-32)
\Text(43,-44)[l]{$k_{134}$}
\LongArrow(-48,32)(-32,48)
\Text(-45,45)[r]{$k_{234}$}
\LongArrow(-48,-32)(-32,-48)
\Text(-48,-30)[br]{$k_1$}
\LongArrow(62,5)(72,5)
\Text(68,9)[b]{$P$}

\end{picture}
\caption{\label{fig:2}
A cut contributing to $H \to b \bar b q \bar q$.
Only the $g_{\rho \sigma}$ term is considered in \eqref{eq:gammar}.} 
\end{figure}
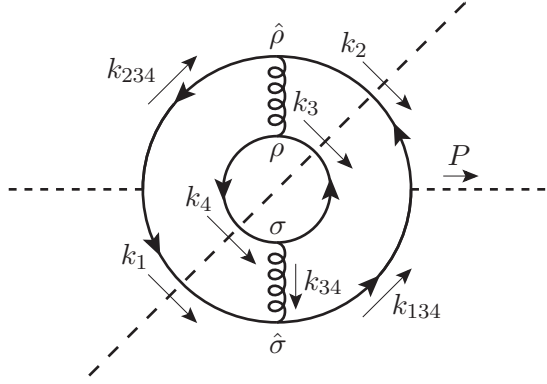

$\tilde \Gamma_V$ develops IR divergences in the form of powers of
$L = \ln(\mu^2/s)$. When splitting the result of the integration in a part which collects all terms containing powers of $L$, dubbed logarithmic part ($L.P.$), plus a remainder, one finds
\bqa
\label{eq:LP}
L.P. \left(
{\cal R}e\left(\tilde \Gamma_V\right)
\right)
= -\frac{s}{64 \pi^5}
\left(L \left(1-\frac{\pi^2}{12}\right) +\frac{L^2}{4}
+\frac{L^3}{24}\right).
\eqa
These logarithms are cancelled by a term contributing to the four-particle cut-diagram in figure~\ref{fig:2}
\bqa
\label{eq:gammar}
\tilde \Gamma_R &=& \frac{1}{(2 \pi)^8}\int d^4{k_1} d^4{k_2} d^4{k_3} d^4{k_4}
\delta_+(k^2_1) \delta_+(k^2_2) \delta_+(\bar k^2_3) \delta_+(\bar k^2_4)
\delta^4(P-k_{1234})\frac{\tilde N_R}{s^2_{34} s_{134}s_{234}}, \nl
\eqa
where the IR behavior is now regulated by the two external massive lines
$k_{3}^2= k_{4}^2= \mu^2$.
$\tilde N_R$ is obtained by applying the GS operation defined
in \eqref{eq:GS} to the denominator of the diagram
\bqa
\label{eq:NR1}
N_R &=& 64 (k_3 \cdot k_4) (k_1 \cdot k_{234})(k_2 \cdot k_{134}).  
\eqa
%
%
%
%
%
the result reads
\bqa
\label{eq:barnr}
\tilde N_R = 8 s_{34}(s-s_{234})(s-s_{134}).
\eqa
$\tilde N_R$ does not depend on $\mu^2$ because in this case there are no contractions of $k_{3,4}$ vectors with the external indices $\hat \rho$ and $\hat \sigma$.

We now can check that the integrand in \eqref{eq:gammar} is the correct object
to cancel the logarithms in \eqref{eq:LP}. In fact, it
can be obtained from \eqref{eq:gammavt} by means of the Cutkosky replacement
in \eqref{eq:cond:1}, together with the relabellings
\bqa
\label{eq:rpl}
q_1 \to k_{34},~~q_2 \to -k_{3},~~p_2 \to k_2,~~p_1 \to -k_{134},
\eqa
inferred by comparing figures~\ref{fig:1} and \ref{fig:2},  and the substitution
\bqa
\frac{1}{(q^2_1+i0^+)^2} \to -\frac{1}{(s_{34}+i0^+)(s_{34}-i0^+)},
\eqa
which is necessary because of the gluon propagator appearing on the r.h.s. of figure~\ref{fig:2}. Therefore, $\tilde \Gamma_R$ must contain a contribution with the same singular behavior of $-\tilde \Gamma_V$. We dub $\tilde \Gamma_{R}^\prime$ such a contribution, and $\tilde N_{R}^\prime$ its numerator function.
The terms proportional to $s_{234}$ or $s_{134}$ in \eqref{eq:barnr} give zero when evaluated at the two-particle cut: they cannot be ``seen'' by $\tilde \Gamma_V$. This leads us to the conclusion that
\bqa
\tilde N_{R}^\prime = 8 s^2 s_{34},
\eqa
which corresponds to ${\tilde {\cal  Z}_V}(q^2_1)$ in \eqref{eq:NVex},
modulo the first replacement in \eqref{eq:rpl}.
Thus, $\tilde \Gamma_{R}^\prime$ is obtained by replacing
$
\tilde N_{R} \to \tilde N_{R}^\prime
$
in \eqref{eq:gammar}. 
An explicit calculation confirms that
\bqa
L.P. \left({\cal R}e\left(\tilde \Gamma_V\right)+\tilde \Gamma_{R}^\prime\right) = 0.
\eqa

\section{Renormalization}
\label{sec:ren}
In this section, we discuss and implement the FDR renormalization program in the context of our calculation.  To do this we need to distinguish, at least conceptually, between UV regulator, IR regulator and renormalization scale. We denote them by $\muuv$, $\muir$ and $\mur$, respectively.  

In the case of IR free observables, the FDR integral operator in \eqref{eq:FDRope} subtracts the UV infinities before integration. For this reason, after taking the asymptotic limit $\muuv \to 0$, $\muuv$ can be directly interpreted as the finite renormalization scale $\mur$ \cite{Pittau:2012zd}. In this sense, FDR directly produces a finite, renormalized result for the loop part: nothing needs to be subtracted from it. However, this result is arbitrary until bare parameters are fixed by experimental measurements. After doing so, if the theory is renormalizable, the scale $\mur$ gets replaced by physical scales, leading to an unambiguous prediction. This can be understood as a finite renormalization necessary to make the theory predictive \cite{Pittau:2013ica}.

In the presence of IR divergences, no distinction is made in the virtual component between $\muuv$ and $\muir$. As a matter of fact, the procedure in section \ref{sec:virt} assumes $\mu= \muuv= \muir$, preventing one from setting
$\mu = \muuv= \mur$, as is possible in the IR free case.
Our solution is fixing the bare parameters in terms of physical quantities before combining virtual and real components. After this is done, the $\mur$ scales get automatically replaced by physical scales, hence the left over $\mu$s are the
$\muir$s which cancel the IR behavior of the real counterpart.

The bare parameters in our calculation are $\alpha_S^0$ and the Yukawa coupling
$y_b^0$. In order to implement our renormalization program we need relations linking them to measured quantities at the appropriate perturbative order, which is one loop for $\alpha_S^0$ and two loops for $y_b^0$.

We are interested in corrections proportional to $N_F$. Hence, $\alpha_S^0$ can be linked to the customary
$\alpha_S^{\mbox{\tiny ${\rm \overline{MS}}$}}(s)$ by using the fact that
the $N_F$ contribution to the running coincides
in FDR and $\overline{\rm MS}$ \cite{Pittau:2013qla}.
As a consequence, we choose our renormalized strong coupling constant to be $\alpha_S= \alpha_S^{\mbox{\tiny ${\rm \overline{MS}}$}}(s)$, that gives the relation
\bqa
\label{eq:a0}
a^0 = a \left(1+a {\delta}^{(1)}_a  \right),
\eqa
with
\bqa
\label{eq:a0anda}
a^0 := \frac{\alpha^0_S}{4 \pi},~~~a := \frac{\alpha_S}{4 \pi},~~~
{\delta}^{(1)}_a= \frac{2}{3} N_F L.
\eqa

The Yukawa coupling is renormalized by using its proportionality to the bottom mass. The corrected bottom propagator at the pole is proportional to
\bqa
\frac{1}{\rlap /p -m^0 + {\Sigma}^{(1)} + {\Sigma}^{(2)}},
\eqa
where $m^0$ is the bare mass and the ${\Sigma}^{(j)}$s are computed in appendix \ref{app:bprop}. This gives a relation between $m^0$ and the pole mass $m$
\bqa
m^0= m+\Sigma^{(1)}+\Sigma^{(2)},
\eqa
which translates into
\bqa
\label{eq:y01}
y^0_b= y_b \left(
1+a^0\delta^{(1)}_y + a^2\delta^{(2)}_y  
\right),
\eqa
with
\bqa
\delta^{(1)}_y &=& -C_F \left(3 L^{\prime \prime} +5  \right), \nl
\delta^{(2)}_y &=& C_F N_F \left({L^{\prime \prime}}^2
+\frac{13}{3}{L^{\prime \prime}}
+ \frac{2}{3} \pi^2+\frac{151}{18}\right).
\eqa
Equation \eqref{eq:y01} contains the bare QCD coupling.
Inserting \eqref{eq:a0} gives the desired two-loop relation between bare and renormalized Yukawa coupling
\bqa
\label{eq:y0}
y^0_b= y_b \left(
1+a \delta^{(1)}_y + a^2
\left(\delta^{(2)}_y+ {\delta}^{(1)}_a \delta^{(1)}_y 
\right)
\right).
\eqa

\section{${ H} \to b \bar b + {jets}$}
\label{sec:H}
In this section, we use FDR to reproduce the physical prediction for the inclusive decay width of the Higgs into two $b$ jets up to the NNLO accuracy in the large $N_F$ limit of QCD. That means computing the observable
\bqa
\label{eq:obs1}
\Gamma^{\mbox{\tiny NNLO}}(y_b)= \Gamma_2^{(0)}(y_b)+ \delta \Gamma^{N_F},
\eqa
where $\Gamma_2^{(0)}(y_b)$ is the tree-level decay width and $\delta \Gamma^{N_F}$ collects all the NNLO terms proportional to $\alpha_S^2N_F$. 

The correction factor $\delta \Gamma^{N_F}$ receives contributions from processes with up to four final-state particles, namely:
\begin{itemize}
\item $H \to b \bar b$ up to two loops;
\item $H \to b \bar b g$ at the tree level;
\item $H \to b \bar b  q \bar q$ at the tree level.
\end{itemize}
The tree- and one-loop two- and three-body decays in the above list contribute to $\delta \Gamma^{N_F}$ through renormalization.
As a matter of fact, due to the scalar nature of the Yukawa coupling,
$\Gamma^{\mbox{\tiny NNLO}}(y_b)$ is a simple process in terms of the contributing tensor structures. Nevertheless, it requires the two-loop renormalization of \eqref{eq:y0}.

In the following, we compute all components in the massless limit of QCD, namely with $m\ne 0$ only in $y_b$. Our notation is as follows.
We dub $V_i^{(j)}$ the $H$ decay amplitudes into $i$ final state partons computed at the $j^{th}$ order of the QCD perturbative expansion. We shall omit for brevity the multiplication of the appropriate quark spinors in any expressions for the $V_i^{(j)}$.
The decay widths are obtained by squaring the amplitudes and are denoted by $\Gamma_i^{(j)}$.

In this paper we focus on the new aspects of FDR at NNLO, namely the procedures presented in sections \ref{sec:virt} and \ref{sec:real}. For this reason we do not go into detail of the calculation of the NLO part.
The corresponding expressions can be computed as described in references \cite{Pittau:2013qla,Gnendiger:2017pys}. However, we emphasize that FDR NLO formulae stay the same also when they contribute to a NNLO calculation. The same holds true for LO expressions multiplying higher order corrections. This is in contrast to $d$-dimensional regularization methods, in which higher powers in the $(d-4)$ expansion must be added.  
 
\subsection{${ H} \to b \bar b$ up to two loops}
\label{sec:hbb2l}
In our conventions, the lowest order ${ H} \to b \bar b$ vertex is
\bqa
V_2^{(0)} = \dkl y_b^0,
\eqa
where $k$ and $l$ are the color indices of the bottom quarks.
Squaring $V_2^{(0)}$ gives the LO ${ H} \to b \bar b$ decay width
\bqa
\label{eq:gamma2}
\Gamma_2^{(0)}(y_b^0)=  (y_b^0)^2 M_H \frac{N_C}{8 \pi}.
\eqa
in which $N_C$ is the number of colors.

\begin{figure}[tbp]
\begin{picture}(200,94)(0,0)
\SetOffset(220,41)
    \SetScale{1}
\Text(-45,1)[r]{$V_2^{(1)}=$}
\DashLine(-40,0)(0,0){3}
\ArrowLine(0,0)(20,20)
\ArrowLine(20,20)(40,40)
\ArrowLine(20,-20)(0,0)
\ArrowLine(40,-40)(20,-20)
\Gluon(20,20)(20,-20){3}{6}
\end{picture}
\caption{\label{fig:hbb1loop}
The one-loop QCD correction to the $Hb \bar b$ vertex.}
\end{figure}
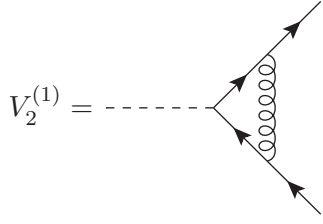

The one-loop correction is depicted in figure \ref{fig:hbb1loop}.
Following reference \cite{Pittau:2013qla} one obtains
\bqa
\label{eq:v21}
V_2^{(1)} = -\dkl\left(\frac{\alpha_S^0}{4 \pi}\right) {y_b^0} C_F {L^\prime}^2,
\eqa
with
\bqa
\label{eq:lprime}
L^\prime := \ln \frac{\mu^2}{-s-i0^+},~~s= M^2_H,~~C_F=\frac{N_C^2-1}{2 N_C}.
\eqa

The globally prescribed integral needed to compute the two-loop correction is given by left part of figure~\ref{fig:1}.
It reads
\bqa
\label{eq:V2H}
\bar V_2^{(2)}= \dkl\frac{y_b}{8} C_F N_F \frac{\alpha_S^2}{\pi^6}
\int [d^4q_1] [d^4q_2]
\frac{\bar N_A+ \bar N_B}{\bar q_1^4 \bar D_1 \bar D_2 \bar q_2^2 \qotsb},
\eqa
with $\bar D_i$ written in \eqref{eq:gammav}.
Here we have replaced bare quantities with renormalized ones, because the difference is ${\cal O}(\alpha^3_S)$.
In the following, we compute $\bar N_{A,B}$ starting form their unbarred counterparts $N_{A,B}$, which can be obtained from \eqref{eq:integrand} by taking
$F_{\hat \rho  \hat \sigma} = \gamma_{\hat \rho} (\rlap/q_1 + \rlap/p_1) (\rlap/q_1 + \rlap/p_2) \gamma_{\hat \sigma}$.
By denoting  $\rlap/\hat q_2 := \gamma_{\hat \rho} q^\rho_2=\gamma_{\hat \sigma} q^\sigma_2$ one finds
\bqa
N_A &=&
  \rlap/q_1 (\rlap/q_1 +\rlap/p_1) (\rlap/q_1 +\rlap/p_2) \rlap/q_2
 +\rlap/q_2 (\rlap/q_1 +\rlap/p_1) (\rlap/q_1 +\rlap/p_2) \rlap/q_1,\nl
N_B &=&
2\rlap/\hat q_2 (\rlap/q_1 +\rlap/p_1) (\rlap/q_1 +\rlap/p_2) \rlap/ \hat q_2
-\frac{N_V}{8s},
\eqa
where the last term originates from the piece we have studied in detail in section \ref{sec:example}, with $N_V$ given in \eqref{eq:NVunb}.
$N_A$ does not contribute to $\bar V_2^{(2)}$. In fact, a $\bar u(p_1)$($u(p_2)$) is understood on the l.h.s.(r.h.s.), so that, by virtue of the Dirac equation, one can replace
$
N_A \to N_A^\prime = D_1  (\rlap/q_1 +\rlap/p_2) \rlap/q_2 +
\rlap/q_2 (\rlap/q_1 +\rlap/p_1)D_2$
, so that
\bqa
N^\prime_A \toGP \bar N^\prime_A= \bar D_1  (\rlap/q_1 +\rlap/p_2) \rlap/q_2 +
\rlap/q_2 (\rlap/q_1 +\rlap/p_1) \bar D_2,
\eqa
which generates scale-less integrals. That explicitly proves the WI in
\eqref{eq:2wi}.
As for the $N_B$ piece, there are several ways \cite{Donati:2013iya,Donati:2013voa} to deal with strings of $\gamma$-matrices to extract the dependence on $(q_i \cdot q_j)$ and $q_i^2$ needed to implement GP. They are based on replacements of the type
$\rlap/q_i \to \rlap /q_i -\mu_i$, 
where the ``masses'' $\mu_i$ serve as a bookkeeping tool.
In this paper we find it more convenient to use Clifford algebra until we reach the configurations~\footnote{Note the invariance of the last term under $\rlap/q_i \to \rlap /q_i -\mu_i$.}
\bqa
\label{eq:slashGP}
\rlap/q_i \rlap/q_i \toGP \bar q^2_i,~~~
\rlap/q_i \rlap/q_j \toGP \frac{1}{2}
\left(\bar q^2_{ij}-\bar q^2_i-\bar q^2_j
+ \rlap/q_i \rlap/q_j-\rlap/q_j \rlap/q_i 
\right).
\eqa
By using this method one finds $N_B \toGP \bar N_B \simeq \bar N^\prime_B$ with
\bqa
\label{eq:NB}
\bar N^\prime_B(\bar q^2_1,\hat q^2_2) =&& -4 (q_1 \cdot p_1) \hat q^2_2
  +8 (q_2 \cdot p_2)\big((q_2 \cdot q_1)+(q_2 \cdot p_1) \big) \nl  
  &&-\frac{8}{s} (P \cdot q_2)\big(
                  (q_1 \cdot p_2)(q_2 \cdot p_1)
                 -(q_1 \cdot p_1)(q_2 \cdot p_2)
                            \big)   
-s \bar q^2_1. 
\eqa
$\bar N^\prime_B$ does not induce the appearance of global UV divergences in 
\eqref{eq:V2H}, hence the numerator function directly reads
$
{\bar {\cal  Z}}(\bar q^2_1,\hat q^2_2)= \bar N^\prime_B(\bar q^2_1,\hat q^2_2)
$.
Thus, the SIC preserving numerator function is
$
{\tilde {\cal  Z}}(q^2_1,q^2_2)= \bar N^\prime_B(q^2_1,q^2_2)
$,
and the two-loop correction is
\bqa
\label{eq:Itilde}
V_2^{(2)}= \dkl\frac{y_b}{8} C_F N_F \frac{\alpha_S^2}{\pi^6}
\int d^4q_1 [d^4q_2]
\frac{{\tilde {\cal  Z}}(q^2_1,q^2_2)}{q_1^4 D_1 D_2 \bar q_2^2 \qotsb}.
\eqa
In terms of the master integrals listed in appendix~\ref{app:C} it reads
\bqa
V_2^{(2)}=  \dkl \frac{4}{3} y_b C_F N_F
\left(\frac{\alpha_S}{4 \pi}\right)^2
{{\tilde I}_1}.
\eqa

\subsection{${ H} \to b \bar b g$ and $H \to b \bar b q \bar q$ at the tree level}
\label{sec:bbqq}

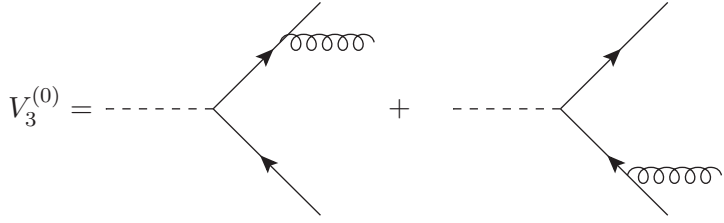
\begin{figure}[tbp]
\begin{picture}(200,94)(0,0)
\SetScale{1}

\SetOffset(160,41)
\Text(-45,1)[r]{$V_3^{(0)}=$}
\DashLine(-40,0)(0,0){3}
\ArrowLine(0,0)(40,40)
\ArrowLine(40,-40)(0,0)
\Gluon(25,25)(60,25){3}{5}
\SetOffset(290,41)
\Text(-45,0)[r]{$+~~~$}
\DashLine(-40,0)(0,0){3}
\ArrowLine(0,0)(40,40)
\ArrowLine(40,-40)(0,0)
\Gluon(25,-25)(60,-25){3}{5}
\end{picture}
\caption{\label{fig:hbbg}
The LO $Hb \bar b g$ vertex.}
\end{figure}

The LO ${ H} \to b \bar b g$ decay width is obtained by squaring the $V_3^{(0)}$
vertex drawn in figure \ref{fig:hbbg} and integrating over a phase-space in which all final-state particles acquire a small mass $\mu$, as described in reference \cite{Pittau:2013qla}.
The result is
\bqa
\label{eq:gamma3}
\Gamma_3^{(0)}= \left(\frac{\alpha_S^0}{4 \pi}\right) \Gamma_2^{(0)}(y^0_b) C_F
\big(
2 L^2 +6L+19-2 \pi^2
\big),
\eqa
where
\bqa
\label{eq:ell}
L := \ln(\mu^2/s).
\eqa

A for $H \to b \bar b q \bar q$, two diagrams contribute to the amplitude $V_4^{(0)}$. They can be read from
figure \ref{fig:hbbg} by allowing the gluon to split into a $q \bar q$ pair.
As described in section \ref{sec:real}, $\Gamma_4^{(0)}$ is obtained by squaring $V_4^{(0)}$ and integrating over a massive 4-particle phase-space $\tilde \Phi_4$ such that $k_1^2= k_2^2= 0$ and  $k_3^2= k_4^2= \mu^2$.
Prior to integration, the integrand should be modified according to the GP and SIC replacements given in \eqref{eq:NRtilde}.
As a result of this, the function to be integrated  is a rational combination of the invariants $s_{34}$, $s_{134}$, $s_{234}$ and $\mu^2$:
\bqa
S(s_{34},s_{134},s_{234},\mu^2),
\eqa
where the $\mu^2$ dependence is induced by \eqref{eq:nogpr}.
It is interesting to note this $\mu^2$ dependence factorizes. In particular, one finds $S(s_{34},s_{134},s_{234},\mu^2)= S^\prime(s_{34},s_{134},s_{234})w(\mu^2)$, with
\bqa
\label{eq:wmu}
w(\mu^2) = \left(1+2 \frac{\mu^2}{s_{34}} \right). 
\eqa
In terms of the integrals reported in appendix~\ref{app:D} the result reads
\bqa
\label{eq:gamma4}
\Gamma_4^{(0)}= \frac{64}{3} C_F N_F \Gamma_2^{(0)}(y_b)
\left(\frac{\alpha_S}{4 \pi}\right)^2
\left(
  {\tilde R}_8
+ {\tilde R}_7
+ {\tilde R}_6
-2{\tilde R}_5
- {\tilde R}_4
\right).
\eqa

\subsection{The large $N_F$ limit of the inclusive width}
Here we gather all the calculated components and compute $\Gamma^{\mbox{\tiny NNLO}}(y_b)$ in \eqref{eq:obs1}. The correction factor
$\delta \Gamma^{N_F}$ receives contributions from processes with up to four partons
\bqa
\label{eq:gammanf}
\delta \Gamma^{N_F}=
\Gamma^{N_F}_2+
\Gamma^{N_F}_3+
\Gamma^{N_F}_4,
\eqa
that are obtained by inserting the renormalization equations \eqref{eq:a0} and \eqref{eq:y0} in the amplitudes given in the previous sections.

One finds
\bqa
\label{eq:g4nf}
\Gamma^{N_F}_2 &=& \Gamma_2^0(y_b)a^2 2 \Re e 
\left(         \delta V^{(2)}_2
+\delta^{(1)}_a \delta V^{(1)}_2
+\delta^{(2)}_y
+ \delta^{(1)}_a\delta^{(1)}_y 
\right), \nl
\Gamma^{N_F}_3&=& a^2 \delta^{(1)}_a \Gamma_2^0(y_b)
C_F
\big(
2 L^2 +6L+19-2 \pi^2
\big), \\
\Gamma^{N_F}_4 &=&
a^2 C_F N_F \Gamma_2^{(0)}(y_b) \frac{4}{9}
\left\{
-L^3
-\frac{19}{2} L^2
-L
\left(
\frac{155}{3}- 2 \pi^2
\right)
+ 30 \zeta_3
+\frac{29}{6}\pi^2-\frac{4345}{36}
\right\}, \nonumber 
\eqa
where
\bqa
\delta V^{(1)}_2=&& -C_F (L^\prime)^2, \nl
\delta V^{(2)}_2=&& \frac{2}{9} C_F N_F
\left(
   {L^\prime}^3
+5 {L^\prime}^2
+  {L^\prime}\left(\frac{56}{3}+\pi^2\right)
   -12 \zeta_3
   +\frac{5}{3}\pi^2
   +\frac{328}{9}
   \right).
\eqa
Equations \eqref{eq:g4nf} are written in a form that highlights 
the contributions generated by renormalization.
Collecting all the pieces gives the IR finite result
\bqa
\label{eq:resg1}
\Gamma^{\mbox{\tiny NNLO}}(y_b)= \Gamma_2^{(0)}(y_b)
\left\{1+
a^2 C_F N_F 
\left(
2\ln^2 \frac{m^2}{s}
-\frac{26}{3} \ln \frac{m^2}{s}
+ 8 \zeta_3 + 2 \pi^2 -\frac{62}{3}
\right)
\right\}.
\eqa

Equation \eqref{eq:resg1} is written in terms of the pole mass $m$.
It is possible to reabsorbe the large logarithms of the ratio $m^2/s$ in
a new Yukawa coupling $y^{\mbox{\tiny ${\rm \overline{MS}}$}}_b$ defined through the known two-loop relation between $m$ and the ${\rm \overline{MS}}$
mass \cite{Bednyakov:2016onn}.
Using the $N_F$ part of it gives
\bqa
\Gamma_2^{(0)}(y_b)= \Gamma_2^{(0)}(y_b^{\mbox{\tiny ${\rm \overline{MS}}$}}(s))
\left\{
1+a^2 C_F N_F
\left(
-2\ln^2 \frac{m^2}{s}+\frac{26}{3}\ln\frac{m^2}{s}
-\frac{4}{3}\pi^2 - \frac{71}{6}
\right)
\right\},
\eqa
hence
\bqa
\label{eq:gammafin}
\Gamma^{\mbox{\tiny NNLO}}(y_b^{\mbox{\tiny ${\rm \overline{MS}}$}}(s))=
\Gamma_2^{(0)}(y_b^{\mbox{\tiny ${\rm \overline{MS}}$}}(s))
\left\{
1+a^2 C_F N_F
\left(8 \zeta_3
+\frac{2}{3}\pi^2 - \frac{65}{2}
\right)
\right\}.
\eqa
Equation \eqref{eq:gammafin} coincides with the known ${\rm \overline{MS}}$ result \cite{Gorishnii:1991zr}.

\section{$\gamma^\ast \to {jets}$}
\label{sec:gamma}
In this section we compute the large $N_F$ limit of the inclusive
$e^+ e^- \to \gamma^\ast \to jets$ production rate up to the NNLO accuracy.
That is the observable
\bqa
\sigma^{\mbox{\tiny NNLO}}= \sigma_2^{(0)}+ \delta \sigma^{N_F},
\eqa
where $\sigma_2^{(0)}$ is the tree-level $e^+ e^- \to \gamma^\ast \to q  \bar q$ cross-section and $\delta \sigma^{N_F}$ contains the QCD corrections proportional to $\alpha_S^2N_F$.
QCD renormalization only involves $\alpha_S$, in this case. Nevertheless, higher rank tensors contribute, so that preserving gauge cancellations and unitarity in such an environment provides a more stringent test for our procedures.
In this respect, $\gamma^\ast \to jets$ is complementary to $H \to b \bar b + jets$.

The processes which contribute to $\delta \sigma^{N_F}$ are
\begin{itemize}
\item{$e^+ e^- \to q  \bar q$ up to two loops};
\item{$e^+ e^- \to q \bar q g$ at the tree level};
\item{$e^+ e^- \to q \bar q q^\prime \bar q^\prime$ at the tree level},
\end{itemize}
where we understand a photon mediating the reactions.
We dub $V^{(j)\beta}_i$ the final-state current producing $i$ partons computed at the $j^{th}$ QCD order, where $\beta$ is the Lorentz index of the virtual photon. The Feynman diagrams representing the vertices are obtained from those in the previous section by replacing the Higgs with a photon. Hence, we do not draw them. Contracting $V^{(j)\beta}_i$ with the initial-state current, squaring and integrating over the phase-space gives the corresponding cross section, denoted by $\sigma^{(j)}_i$.

In the following, we describe the FDR computation of the various components.
\subsection{$e^+ e^- \to q  \bar q$ up to two loops}
The lowest order vertex is
\bqa
V^{(0)\beta}_2= -i e Q_q \dkl \gamma^\beta,
\eqa
where $Q_q$ is the electric charge of the quark. The corresponding cross section reads
\bqa
\sigma^{(0)}_2= N_C \frac{4}{3} \pi \frac{\alpha^2}{s} Q_q^2,
\eqa
in which $\alpha$ is the fine-structure constant.

The computation of $V^{(1)\beta}_2$ is described in \cite{Gnendiger:2017pys}.
The result is
\bqa
\label{eq:v12be}
V^{(1)\beta}_2= - a^0 C_F V^{(0)\beta}_2
\left({L^\prime}^2+3 L^\prime+7 \right),~~
\eqa
with $a^0$ and $L^\prime$ defined in \eqref{eq:a0anda} and \eqref{eq:lprime}, respectively.

The globally prescribed two-loop integral we need to compute $V_2^{(2)\beta}$ reads
\bqa
\label{eq:V2gammab}
\bar V_2^{(2)\beta}= (-i e Q_q \dkl)\frac{C_F N_F}{8} \frac{\alpha_S^2}{\pi^6}
\int [d^4q_1] [d^4q_2]
\frac{\bar N^\beta}{\bar q_1^4 \bar D_1 \bar D_2 \bar q_2^2 \qotsb}.
\eqa
The unbarred $N^\beta$ is obtained from \eqref{eq:integrand} with
$F_{\hat \rho  \hat \sigma} = \gamma_{\hat \rho} (\rlap/q_1 + \rlap/p_1) \gamma^\beta (\rlap/q_1 + \rlap/p_2) \gamma_{\hat \sigma}$ by using the fact that, according to the WI in \eqref{eq:2wi}, the term proportional to $q^\rho_2 q^\sigma_1+q^\rho_1 q^\sigma _2$ in the fermion trace does not contribute~\footnote{The proof is analogous to the one given in section \ref{sec:hbb2l}.}. Hence
\bqa
N^\beta= 2 (q_2 \cdot \qot)(\rlap/q_1 +\rlap/p_2)\gamma^\beta(\rlap/q_1 +\rlap/p_1)
      +2\rlap/\hat q_2 (\rlap/q_1 +\rlap/p_1)\gamma^\beta(\rlap/q_1 +\rlap/p_2) \rlap/ \hat q_2.
\eqa
Using tensor decomposition and the $p_1 \leftrightarrow p_2$ symmetry gives
$N^\beta \toGP \bar N^\beta \simeq \bar M^\beta$ with
\bqa
\bar M^\beta(\bar q_1^2,\hat q^2_2)=&&+2 \gamma^\beta
\bigg\{
4 (q_2 \cdot p_1)  (q_2 \cdot p_2) 
+\frac{4}{s} (q_2 \cdot P) 
\big(
 (q_1 \cdot p_1) (q_2 \cdot p_2)
-(q_1 \cdot p_2) (q_2 \cdot p_1) 
\big)
\nl
&&
+ 2 \bar q^2_1  (q_1 \cdot p_1) -\frac{s}{2} \bar q^2_1
\bigg\} +8 (q_1 \cdot p_1) \rlap/q_2 \gamma^\beta \nl
&&-4 \hat q^2_2
\bigg\{
(q_1 \cdot p_1) \gamma^\beta
+ q_1^\beta \rlap/q_1 \frac{-2(q_1 \cdot p_1)}{\bar q^2_1}
\bigg\},
\eqa
where the factor $-2(q_1 \cdot p_1)/\bar q^2_1$ multiplying the last term subtracts its GV.
Thus, the SIC preserving numerator function is
$
{\tilde {\cal  Z}}^\beta(q^2_1,q^2_2)= \bar M^\beta(q_1^2,q^2_2) 
$, giving
\bqa
\label{eq:V2gamma}
V_2^{(2)\beta}= (-i e Q_q \dkl) \frac{C_F N_F}{8} \frac{\alpha_S^2}{\pi^6}
\int d^4q_1 [d^4q_2]
\frac{{\tilde {\cal  Z}^\beta}(q^2_1,q^2_2)}{q_1^4 D_1 D_2 \bar q_2^2 \qotsb}.
\eqa
In terms of the two-loop integrals in appendix~\ref{app:C} one finds
\bqa
\label{eq:V2gammaI}
V_2^{(2)\beta}= V_2^{(0)\beta}\frac{16 C_F N_F}{3} a^2
\left({\tilde I}_3 -{\tilde I}_2 +\frac{{\tilde I}_1}{4}
\right).
\eqa

\subsection{$e^+ e^- \to q \bar q g$ and $e^+ e^- \to q \bar q q^\prime \bar q^\prime$ at the tree level}
A NLO computation produces
\bqa
\label{eq:sigma3}
\sigma_3^{(0)}= a^{0} \sigma_2^{(0)} C_F
\big(
2 L^2 +6L+17-2 \pi^2
\big),
\eqa
with $L$ is given in \eqref{eq:ell}.

As for $\sigma_4^{(0)}$, it is obtained
by computing the amplitude squared, modifying it
according to the prescription  in \eqref{eq:NRtilde}
and integrating over the $\tilde \Phi_4$ phase-space.
In terms of the integrals in appendix~\ref{app:D} the result reads
\bqa
\label{eq:sigma4}
\sigma_4^{(0)}= \frac{64}{3} C_F N_F \sigma_2^{(0)} a^2
\left(
   {\tilde R}_7
+  {\tilde R}_6
-2 {\tilde R}_5
-  {\tilde R}_4
+  {\tilde R}_3
+2 {\tilde R}_2
-2 {\tilde R}_1
\right).
\eqa

\subsection{The large $N_F$ limit of the inclusive jet production rate}
Here we collect all components needed to compute $\sigma^{\mbox{\tiny NNLO}}$.
The correction can be split as follows
\bqa
\label{eq:sigmanf}
\delta \sigma^{N_F}=
\sigma^{N_F}_2+
\sigma^{N_F}_3+
\sigma^{N_F}_4,
\eqa
where the various contributions are obtained by inserting
\eqref{eq:a0} in the results of the previous sections. One has
\bqa
\label{eq:sigma4p}
\sigma^{N_F}_2  &=&  \sigma^{(0)}_2 a^2 2 \Re e 
\left(\delta \hat V^{(2)}_2
+\delta^{(1)}_a \delta \hat V^{(1)}_2
\right),\nl
\sigma_3^{N_F}  &=& a^2 \delta^{(1)}_a\sigma_2^{(0)} C_F
\big(
2 L^2 +6L+17-2 \pi^2 \big), \\
\sigma_4^{N_F}  &=& a^2 C_F N_F \sigma_2^{(0)} \frac{4}{9}
\left\{
-L^3-\frac{19}{2}L^2 -L\left( \frac{146}{3}-2 \pi^2\right)
+30 \zeta_3 + \frac{19}{3}\pi^2-\frac{2123}{18}
\right\}, \nonumber
\eqa
with
\bqa
\delta \hat V^{(1)}_2 =&& -C_F
\left(
{L^\prime}^2+3 L^\prime+7 
\right), \nl
\delta \hat V^{(2)}_2=&& \frac{2}{9} C_F N_F
\left(
   {L^\prime}^3
+\frac{19}{2} {L^\prime}^2
+  {L^\prime}\left(\frac{265}{6}+\pi^2\right)
   -12 \zeta_3
   +\frac{19}{6}\pi^2
   +\frac{3355}{36}
   \right).
\eqa
Gathering all the pieces gives
\bqa
\sigma^{\mbox{\tiny NNLO}}= \sigma_2^{(0)}
\left\{1+
 a^2 C_F N_F 
\left(
8 \zeta_3 -11
\right)
\right\}, 
\eqa
which reproduces the ${\rm \overline{MS}}$ result
\cite{GehrmannDeRidder:2004tv}.

\section{Conclusion and outlook}
\label{sec:conc}
In this paper we have demonstrated that a fully four-dimensional framework
to compute NNLO quark-pair corrections can be constructed based on the requirement of preserving the two principles given in \eqref{eq:prin}.
The FDR idea of enforcing gauge invariance and unitarity at the level of the UV subtracted integrands is at the base of the procedures we have used to define UV and IR divergent integrals.

A few advantages of such an approach that have appeared in our calculation are, for the UV part
\begin{itemize}
\item no (explicit or implicit) UV counterterms have to be included in the Lagrangian;
\item lower-order substructures are used in higher-order calculations without any modification (see e.g. \eqref{eq:v21} and \eqref{eq:v12be});
\item  renormalization is equivalent to the process of expressing (finite) bare parameters in terms of measurable observables (e.g. \eqref{eq:a0} and \eqref{eq:y0}).
\end{itemize}
For the IR sector
\begin{itemize}
\item infrared divergences in the real component directly show up in terms of logarithms of a small
  cut-off parameter $\muir$, with no need for a prior subtraction of $1/(d-4)$ poles (see, for instance, the four-parton rates in \eqref{eq:g4nf} and \eqref{eq:sigma4p});
\item one-to-one integrand correspondences can be written down between virtual and real
 contributions (see section \ref{sec:example}).
\end{itemize}

In this paper we have focused our attention on a special class of NNLO corrections. However, we believe that the basic principles that have guided us towards a consistent treatment of all the pieces contributing to the final NNLO answer will remain valid also when considering more complicated environments, with the final aim of constructing a completely general procedure including also initial state IR singularities. This is certainly the main subject of our future investigations.
Other possible directions are: using $\muir$ as a separation parameter in slicing-based subtraction methods at NNLO \cite{Boughezal:2015ded}, or exploiting the virtual/real integrand correspondence to construct four-dimensional local counter-terms directly from the virtuals.~\footnote{A DReg algorithm along these lines has been recently proposed in \cite{Magnea:2018ebr}.}
On a more general ground, we envisage that the intrinsic four-dimensionality
of FDR can pave the way to new numerical methods and that there is room for fully exploiting its potential in NNLO calculations.

\appendix

\section{Sub-integration consistency with and w/o IR
  divergences}
\label{app:A}
When no IR infinities are present, the mismatch between equations
\eqref{eq:inc1} and \eqref{eq:inc2} is cured by adding the so called {\em extra-extra integrals} (EEI) introduced in \cite{Page:2015zca}. Their exact definition is not needed here. It suffices to say that terms proportional to the difference
\bqa
q^2_2 -\bar q^2_2 = \mu^2 
\eqa
are included. They multiply UV $1/\mu^2$ poles and generate logarithms of $\mu^2$ that restore the correct renormalization properties of the two-loop amplitude.
Such contributions are missed by \eqref{eq:inc2}. 

In the presence of IR divergences an additional complication is generated by the GP $q^2_1 \to \bar q^2_1$ in \eqref{eq:inc1} and \eqref{eq:inc2}. After GV subtraction, the difference
\bqa
\mu^2= q^2_1 -\bar q^2_1
\eqa
also hits $1/\mu^2$ poles of IR origin. This gives rise to 
different renormalization constants for processes with or without IR divergences, which is unacceptable. This leads to the choice of letting $q^2_1$ unbarred, as discussed in section \ref{sec:virt}. For the sake of consistency, also the EEIs part needs to be modified accordingly. The problem is that the EEIs become unregulated when unbarring $\bar q^2_1$ at the integrand level. 
The solution to this is replacing EEIs with the difference of two ordinary FDR integrals, generated by the combination
\bqa
q^2_2 -\bar q^2_2,
\eqa
which is sometimes referred as an {\em extra-integral} (EI).
One shows that EEIs and EIs share the same logarithmic content, which fixes the correct UV behavior. In addition, EIs admit the $\bar q^2_1 \to q^2_1$ limit that matches the rest of the calculation.

In summary, the solution presented in section \ref{sec:virt} is equivalent to the following procedure:
\begin{itemize}
\item apply GP;  
\item subtract GV;
\item downgrade $\bar q^2_1 \to q^2_1$ in the result;  
\item identify the EEIs to be added (using the same algorithm as in the IR-free case);    
\item replace each EEI with the corresponding EI.
\end{itemize}
It would be interesting to establish whether this strategy works also for IR finite two-loop calculations. That would make unnecessary the use of the EEIs. We leave this to further investigations.

\section{Massless wave-function corrections}
\label{app:B}
Wave function corrections are generated when the lower gluon in figure~\ref{fig:BVR}-(b) reconnects to the emitting massless parton. In this appendix, we use the results of section~\ref{sec:virt} to demonstrate that they vanish.

The relevant integrand is obtained by taking
$F_{\hat \rho  \hat \sigma} = \gamma_{\hat \rho} (\rlap/q_1 + \rlap/p) \gamma_{\hat \sigma}$
in \eqref{eq:integrand}, that gives $N   = N_A+N_B$ with
\bqa
\label{eq:NAB}
N_A =
  \rlap/q_1 (\rlap/q_1 +\rlap/p) \rlap/q_2
 +\rlap/q_2 (\rlap/q_1 +\rlap/p) \rlap/q_1~~{\rm and}~~
N_B =
2\rlap/\hat q_2 (\rlap/q_1 +\rlap/p) \rlap/ \hat q_2
+2(q_2 \cdot \qot)(\rlap/q_1 +\rlap/p).
\eqa
Furthermore, $D = q^4_1 D_p$, so that the integrals we have to consider are 
\bqa
\bar I_{A,B} = \int [d^4q_1] [d^4q_2] \frac{\bar N_{A,B}}{\bar q^4_1 \bar D_p \bar q^2_2 \qotsb}.
\eqa
One finds
\bqa
N_A \toGP \bar N_A = 2 \rlap/ q_2 \bar D_p -\rlap/p (\bar q^2_{12}-\bar q^2_1-\bar q^2_2)
-\frac{1}{2} \rlap/p
\left(\rlap/q_1 \rlap/q_2-\rlap/q_2 \rlap/q_1
\right)
-\frac{1}{2} 
\left(\rlap/q_2 \rlap/q_1-\rlap/q_1 \rlap/q_2
\right) \rlap/p.
\eqa
Only the third term contributes to $\bar I_{A}$. All the others generate vacua or result from contractions of antisymmetric combinations of
$\gamma$-matrices with symmetric integrals. Thus
\bqa
\bar I_{A} = \rlap /p \int [d^4q_1] [d^4q_2] \frac{1}{\bar q^2_1 \bar D_p \bar q^2_2 \qotsb}.
\eqa
$\bar I_{A}$ only depends on $p^2= 0$. In addition, it is both UV divergent and logarithmically IR divergent, so that it is a {\em scale-less} integral. Such integrals vanish in FDR as a consequence of an exact cancellation between UV and IR singularities, therefore $\bar I_{A}=0$, as required by \eqref{eq:2wi}.
In the same way, $\bar I_{B}$ is fully scale-less
\bqa
\label{eq:NWF}
\bar I_{B}=
\int [d^4q_1] [d^4q_2]
\frac{ -2 \hat q^2_2 (\rlap/q_1+\rlap/p)
-\bar q^2_1 (\rlap/q_1+\rlap/p)
-2 \bar q^2_1 \rlap/q_2+4(q_2 \cdot p)
\rlap/ q_2}{\bar q^4_1 \bar D_p \bar q^2_2 \qotsb} = 0,
\eqa
so that self-energy corrections $\bar I_{A} +\bar I_{B}$ vanish.

The proof that scale-less integrals do not contribute can be found in \cite{Donati:2013voa}. Here we prove that they vanish also when $q^2_1$ is unbarred, as in \eqref{eq:tildeI}.
We concentrate on first term of \eqref{eq:NWF}
\bqa
\bar I_{C} :=
\int [d^4q_1] [d^4q_2]
\frac{\bar N_C }
{\bar q^4_1 \bar D_p \bar q^2_2 \qotsb},~~~ \bar N_C := \hat q^2_2 (\rlap/q_1+\rlap/p).
\eqa
The proof is unchanged for all the other contributions.
A GV subtraction is needed in front of $\hat q^2_2 \rlap/q_1$. This is
achieved by using twice the identity in \eqref{eq:ident}
\bqa
\label{eq:GVS2}
\frac{1}{\bar D_p}= \left[\frac{1}{\bar q_1^2}
-\frac{2 (q_1 \cdot p)}{\bar q_1^4}\right]_{V} 
+4\frac{(q_1 \cdot p)^2}{\bar q_1^4 \bar D_p}. 
\eqa
The $\hat q^2_2 \rlap/p$ piece is less UV divergent, so that a single subtraction
is sufficient
\bqa
\label{eq:GVS1}
\frac{1}{\bar D_p}= \left[\frac{1}{\bar q_1^2}\right]_{V}
-\frac{2 (q_1 \cdot p)}{\bar q_1^2 \bar D_p}.
\eqa
The vacua are subtracted by the integral operator. That defines the numerator function associated with $\bar N_C$
\bqa
{\bar {\cal  Z}_C}(\bar q^2_1,\hat q^2_2)=
4(\hat q^2_2 \rlap/q_1)\frac{(q_1 \cdot p)^2}{\bar q_1^4} 
-2(\hat q^2_2 \rlap/p)\frac{(q_1 \cdot p)}{\bar q_1^2}.
\eqa
Hence
\bqa
{\bar {\cal  Z}_C}(\bar q^2_1,\hat q^2_2) \toSIC
{\tilde {\cal  Z}_C}(q^2_1,q^2_2)= 
4(q^2_2 \rlap/q_1)\frac{(q_1 \cdot p)^2}{q_1^4} 
-2(q^2_2 \rlap/p)\frac{(q_1 \cdot p)}{q_1^2},
\eqa
which produces
\bqa
\tilde I_{C}=
4 \int d^4q_1 [d^4q_2] 
\frac{q^2_2 \rlap/q_1(q_1 \cdot p)^2}{q^8_1 D_p \bar q^2_2 \qotsb} 
-2 \rlap/p \int d^4q_1 [d^4q_2]
\frac{q^2_2 (q_1 \cdot p)}{q^6_1 D_p \bar q^2_2 \qotsb}.
\eqa
$\tilde I_{C}$ diverges logarithmically in the double collinear configuration in the absence of regulator. The barred $q_2$-type denominators are sufficient to regulate this. That is a consequence of the fact that \eqref{eq:GVS2} and \eqref{eq:GVS1} do not alter the IR power counting.
By tensor decomposition $\tilde I_{C} \sim \rlap /p \left(p^2/\mu^2 + {\cal O}(p^4/\mu^4)\right)$, so that  it vanishes on-shell.
In summary, the GV subtraction  does not leave finite pieces in scale-less integrals.

\section{Correcting the bottom propagator}
\label{app:bprop}

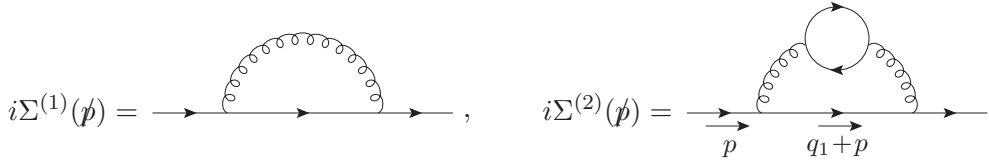
\begin{figure}[tbp]
\begin{picture}(200,64)(0,0)
    
\SetScale{0.7}

\SetOffset(140,15)
\Text(-61.6,1)[r]{$i\Sigma^{(1)}(\rlap /p)=$}
\GlueArc(0,0)(40,0,180){3}{14}
\ArrowLine(-80,0)(-40,0)
\ArrowLine(-40,0)(40,0)
\ArrowLine(40,0)(80,0)
\Text(60,-2)[l]{,}
\SetOffset(340,15)
\Text(-61.6,1)[r]{$i\Sigma^{(2)}(\rlap /p)=$}
\GlueArc(0,0)(40,0,65){3}{5}
\GlueArc(0,0)(40,115,180){3}{5}
\ArrowArc(0,40)(-17,180,0)
\ArrowArc(0,40)(-17,0,-180)
\ArrowLine(-80,0)(-40,0)
\ArrowLine(-40,0)(40,0)
\ArrowLine(40,0)(80,0)
\LongArrow(-70,-7)(-50,-7)
\LongArrow(-10,-7)(10,-7)
\Text(-40.2,-10.4)[t]{\small $p$}
\Text(0,-8.4)[t]{\small $q_1\!+\!p$}
\end{picture}
\caption{\label{fig:sigma12}
The one- and two-loop QCD corrections to the massive bottom propagator. $N_F$ quarks run in the loop.}
\end{figure}

To renormalize the Yukawa coupling, we need the one- and two-loop QCD corrections of figure \ref{fig:sigma12} computed at the value $\rlap /p =m$. We dub them $\Sigma^{(j)}:= \Sigma^{(j)}(\rlap /p =m)$.

One finds
\bqa
\label{eq:sigma1}
{\Sigma}^{(1)} = -m \left(\frac{\alpha^0_S}{4 \pi}\right)
C_F \left(3 L^{\prime \prime} +5  \right),
\eqa
with
\bqa
\label{eq:lprimeprime}
{L^{\prime \prime}}:= \ln \frac{\mu^2}{m^2}.
\eqa

As for the second order contribution, one has
\bqa
\label{eq:sigma2a}
i \bar{\Sigma}^{(2)}= -i \frac{2}{\pi^4}  C_F N_F 
\left(\frac{\alpha_S}{4 \pi}\right)^2 
\int [d^4q_1] [d^4q_2]
\left.\frac{\bar N(\rlap /p= m)}{\bar q_1^4 \bar D_p \bar q_2^2 \qotsb}
\right|_{p^2= m^2},
\eqa
with
$
D_p= q_1^2+2 (q_1 \cdot p)$. 
The unbarred $N(\rlap /p)$ is given by \eqref{eq:integrand} with
$
F_{\hat \rho  \hat \sigma} = \gamma_{\hat \rho} (\rlap/q_1 + \rlap/p +m) \gamma_{\hat \sigma}
$.
The result reads $N(\rlap /p) = N_A(\rlap /p)+N_B(\rlap /p)$, where
\bqa
N_A(\rlap /p) =&&
 \rlap/q_1(\rlap/q_1 + \rlap/p+m) \rlap/q_2
 +\rlap/q_2(\rlap/q_1 + \rlap/p+m) \rlap/q_1 
 = 2 \rlap/q_2 D_p
  - (\rlap/p - m) \rlap/q_1 \rlap/q_2
  - \rlap/q_2 \rlap/q_1 (\rlap/p - m), \nl 
N_B(\rlap /p) =&&
2 (q_2 \cdot \qot)
(\rlap/q_1 + \rlap/p- 2m)
                   +2 \rlap/\hat q_2 (\rlap/q_1 + \rlap/p + m)\rlap/\hat q_2 .
\eqa
When barring $N_A$ one obtains a vanishing contribution to $i \bar{\Sigma}^{(2)}$.
As for $N_B$, one computes
$
N_B(\rlap /p= m)= 2 (q_2 \cdot \qot)(\rlap/q_1 -m)
+4(q_1 \cdot q_2)\rlap/q_2
-2 \hat q^2_2  (\rlap/q_1 -2m)$.
Using tensor decomposition gives
\bqa
N_B(\rlap /p= m) \toGP \bar N_B(\rlap /p= m) \simeq
\bar N^\prime_B(\rlap /p= m),
\eqa
with
\bqa
\bar N^\prime_B(\rlap /p= m)= m
\Big(
\bar q^2_1 -2 \hat q^2_2
\big((q_1 \cdot p)/m^2-2
\big)
\Big).
\eqa
%
%
To subtract the GV from $\bar N^\prime_B$ we expand $1/\bar D_p= 1/\bar q^2_1+\bar f/\bar D_p$ with
$\bar f= -2(q_1 \cdot p)/\bar q^2_1$, that gives the numerator function
\bqa
    {\bar {\cal  Z}}(\bar q^2_1,\hat q^2_2)=
m \Big(
\bar q^2_1\bar f -2 \hat q^2_2
\left({(q_1 \cdot p)}/{m^2}\bar f^2-2\bar f
\right)
\Big).
\eqa
Hence
\bqa
{\bar {\cal  Z}}(\bar q^2_1,\hat q^2_2) \toSIC
{\tilde {\cal  Z}}(q^2_1,q^2_2)=
m \Big(
q^2_1 f -2 q^2_2
\left({(q_1 \cdot p)}/{m^2}f^2-2 f
\right)
\Big),~f= -2\frac{(q_1 \cdot p)}{q^2_1}.
\eqa
In summary, the two-loop correction is
\bqa
\label{eq:sigma2}
 {\Sigma}^{(2)}=&& - \frac{2}{\pi^4}  C_F N_F 
\left(\frac{\alpha_S}{4 \pi}\right)^2 
\int d^4q_1 [d^4q_2]
\left.\frac{{\tilde {\cal  Z}}(q^2_1,q^2_2)}{q_1^4 D_p \bar q_2^2 \qotsb}
\right|_{p^2= m^2} \nl
=&& 4 m \left(\frac{\alpha_S}{4 \pi}\right)^2   C_F N_F\left(2 {\tilde I_5}-{\tilde I_4}\right),
\eqa 
with  ${\tilde I_{4,5}}$ written in appendix \ref{app:C}.

\section{The virtual master integrals}
\label{app:C}
In this appendix we sketch out the computation of the two-loop integrals appearing in our calculation.

The $q_2$ integration is performed first.~\footnote{Assuming the appropriate GV subtraction in the rest of the integral.} This means computing
\bqa
B^{\,;\alpha;\alpha \beta }:=
\int [d^4q_2]
\frac{1; q_2^\alpha; q_2^\alpha q_2^\beta}{\bar q^2_2 \qotsb}.
\eqa
As for $B$, we use the expansion in \eqref{eq:svss}
to subtract its sub-vacuum. Then we use Feynman parametrization and integrate over the UV finite remainder. The result is
\bqa
B= -i \pi^2 q^2_1 \int_0^1 dx
\left(\frac{1}{x}-2
\right)\frac{1}{D_0},
\eqa
with
\bqa
\label{eq:D0}
D_0 := q^2_1 -\mu^2_0,~~~\mu^2_0 := \frac{\mu^2}{x(1-x)}. 
\eqa
To determine $B^{\alpha}$ we use tensor decomposition
\bqa
B^{\alpha}=
\frac{q_1^\alpha}{2}
\int [d^4q_2]
\frac{\qots-q^2_2-q^2_1 }{q^2_1\bar q^2_2 \qotsb}.
\eqa
The first two terms cancel each other due to the $q^2_2 \leftrightarrow \qots$
symmetry of the integral. Thus
\bqa
\label{eq:rank1}
B^{\alpha}= -
\frac{q_1^\alpha}{2} B.
\eqa
Finally, tensor decomposition gives
\bqa
B^{\alpha\beta}=
\frac{1}{3}
 \int [d^4q_2]\frac{1}{\bar q^2_2 \qotsb}
  \left\{
  q^2_2
        \left(
        g^{\alpha\beta}-\frac{q_1^\alpha q_1^\beta}{q^2_1}
        \right)
-\frac{(q_1 \cdot q_2 )^2}{q^2_1}        
        \left(
        g^{\alpha\beta}-4 \frac{q_1^\alpha q_1^\beta}{q^2_1}
        \right)
  \right\}.
\eqa
The coefficients are obtained by subtracting the SV
by means of \eqref{eq:subsubt}, and integrating over the finite part. The result reads
\bqa
\int [d^4q_2]\frac{q_2^2}{\bar q^2_2 \qotsb}= 
-\frac{i \pi^2}{2} q^4_1 \int_0^1 dx
\left(4x^2-1
\right)\frac{1}{D_0},~~~
\int [d^4q_2]\frac{(q_1 \cdot q_2)^2}{\bar q^2_2 \qotsb}= \frac{q^4_1}{4}  B.
\eqa

When inserting these results in equations \eqref{eq:Itilde} and \eqref{eq:V2gamma} one finds that the two-loop vertex corrections can be  expressed in terms of three master integrals
\bqa
    {\tilde I}_j := s^{2-j} \frac{i}{\pi^2}  \int_0^1 dx \left(
\frac{1}{x}-3+4 x^2
\right)
 \int d^4q_1
    \frac{(q_1 \cdot p_1)^{j-1}}{D_0 D_1 D_2},~~~{j= 1,2,3},  
\eqa
where ${\tilde I}_2$ is UV finite because $p_1^2= 0$. 
Integrating over $q_1$ and $x$ and neglecting ${\cal O} (\mu^2)$ terms gives
\bqa
{\tilde I}_1 &=& \frac{1}{6}
\left(
   {L^\prime}^3
+5 {L^\prime}^2
+  {L^\prime}\left(\frac{56}{3}+\pi^2\right)
   -12 \zeta_3
   +\frac{5}{3}\pi^2
   +\frac{328}{9}
   \right), \nl
{\tilde I}_2 &=& -\frac{1}{4}
\left(
              {L^\prime}^2
+ \frac{16}{3}{L^\prime} 
+\frac{\pi^2}{3}
+\frac{104}{9}
\right),\nl
{\tilde I}_3 &=&-\frac{1}{16}
\left(
              {L^\prime}^2
+ \frac{13}{3}{L^\prime} 
+\frac{\pi^2}{3}
+\frac{151}{18}
\right),
\eqa
with $L^\prime$ given in \eqref{eq:lprime}.

Finally, the two-loop integrals in \eqref{eq:sigma2} are
\bqa
    {\tilde I}_4 &:=& \frac{i}{\pi^2}  \int_0^1 dx \left(
\frac{1}{x}-4+8 x^2
\right)
\left. \int d^4q_1
 \frac{(q_1 \cdot p_1)}{q_1^2 D_p D_0}\right|_{p^2= m^2}
 \nl
    {\tilde I}_5 &:=& \frac{i}{\pi^2m^2}  \int_0^1 dx \left(
1-4 x^2
\right)
 \left.\int d^4q_1
    \frac{(q_1 \cdot p_1)^3}{q_1^4 D_p D_0}\right|_{p^2= m^2}.  
\eqa
Their asymptotic expansions read
\bqa
{\tilde I}_4 =
-\frac{1}{4} {L^{\prime \prime}}^2
-\frac{7}{6} {L^{\prime \prime}}
-\frac{\pi^2}{6}
-\frac{41}{18}~~~{\rm and}~~~
{\tilde I}_5 =
-\frac{1}{24} {L^{\prime \prime}}
-\frac{13}{144}, 
\eqa
with ${L^{\prime \prime}}$ in \eqref{eq:lprimeprime}.

\section{The real integrals}
\label{app:D}
The real component of the NNLO corrections computed in this paper can be expressed  in terms
of the following eight integrals
\bqa
&&{\tilde R_1} := \frac{1}{s \pi^3} \int {d^4 \tilde \Phi_4}
\,w(\mu^2)\frac{1}{s_{134}},  \nl
&&{\tilde R_2} := \frac{1}{\pi^3} \int {d^4 \tilde \Phi_4}
\,w(\mu^2)\frac{1}{s_{134} s_{234}},  \nl
&&{\tilde R_3} :=\frac{1}{s \pi^3}   \int {d^4 \tilde \Phi_4}
\,w(\mu^2)\frac{s_{34}}{s_{134} s_{234}},  \nl
&&{\tilde R_4} := \frac{1}{\pi^3} \int {d^4 \tilde \Phi_4}
\,w(\mu^2)\frac{1}{s^2_{134}},  \nl
&&{\tilde R_5} := \frac{1}{\pi^3} \int {d^4 \tilde \Phi_4}
\,w(\mu^2)\frac{1}{s_{134} s_{34}}, \nl
&&{\tilde R_6} := \frac{1}{s \pi^3}  \int {d^4 \tilde \Phi_4}
\,w(\mu^2)\frac{s_{234}}{s_{134} s_{34}}, \nl
&&{\tilde R_7} := \frac{s}{\pi^3}  \int {d^4 \tilde \Phi_4}
\,w(\mu^2)\frac{1}{s_{34} s_{134} s_{234}}, \nl
&&{\tilde R_8} := \frac{1}{s \pi^3} \int {d^4 \tilde \Phi_4}
\,w(\mu^2)\frac{1}{s_{34}},
\eqa
with $w(\mu^2)$ given in \eqref{eq:wmu}.

To compute the ${\tilde R}_i$s  it is convenient to use the following phase-space parametrization
\bqa
\label{eq:pspar}
\int d \tilde \Phi_4= \frac{s^2\pi^3}{8}
\int _{4 \epsilon^2}^{(1-2\epsilon)^2} \!\!\!\!\!\!dz\, \sqrt{1-4 \epsilon^2/z}
\int_z^1 dy
\int _{\frac{z}{y}}^{1-y+z} dx ,
\eqa
where
$
z = {s_{34}}/{s}
$,
$
y = {s_{234}}/{s}
$,
$
x = {s_{134}}/{s}
$,
and 
$
\epsilon^2 = {\mu^2}/{s}
$.
The asymptotic $\mu^2 \to 0$ behavior can be extracted with the change of variable
$
w= 4\epsilon^2/z,
$
to be used when $\lim_{\epsilon \to 0}$ cannot be taken before integration.

The first three integrals are IR finite
\bqa
    {\tilde R_1}&=& \frac{1}{32},\nl
    {\tilde R_2}&=& \frac{\pi^2}{48} -\frac{1}{8},\nl
    {\tilde R_3}&=& -\frac{\pi^2}{96}+\frac{7}{64}.
\eqa
As for the remaining ones, \eqref{eq:pspar} gives
\bqa
\label{eq:R4to8}
    {\tilde R_4}&=& -\frac{L}{16}-\frac{25}{96} ,\nl
    {\tilde R_5}&=&  \frac{L^2}{16}+\frac{11}{24}L
                     -\frac{\pi^2}{48}+\frac{85}{72},\nl
    {\tilde R_6}&=&  \frac{L^2}{32}+\frac{11}{48}L
                     -\frac{\pi^2}{96}+\frac{349}{576} ,\nl
    {\tilde R_7}&=&  -\frac{L^3}{48}-\frac{5}{48}L^2
    +L \left(\frac{\pi^2}{24}-\frac{7}{18}\right)
    +\frac{5}{8} \zeta_3+\frac{5}{72}\pi^2-\frac{41}{54},\nl
    {\tilde R_8}&=& {\tilde R_4},
\eqa
with $L$ written in \eqref{eq:ell}.

\acknowledgments
R.P. acknowledges the financial support of the MINECO project FPA2016-78220-C3-3-P and the hospitality of the CERN TH department during the completion of this work. Our figures are prepared with {\tt Axodraw} \cite{Vermaseren:1994je}.

\bibliography{paper.bib}

\providecommand{\href}[2]{#2}\begingroup\raggedright\begin{thebibliography}{10}

\bibitem{Sterman:1994ce}
G.~F. Sterman, \emph{An Introduction to quantum field theory}. Cambridge
  University Press, 1994.

\bibitem{Cutkosky:1960sp}
R.~E. Cutkosky, \emph{{Singularities and discontinuities of Feynman
  amplitudes}}, \href{https://doi.org/10.1063/1.1703676}{\emph{J. Math. Phys.}
  {\bfseries 1} (1960) 429}.

\bibitem{Veltman:1963th}
M.~J.~G. Veltman, \emph{{Unitarity and causality in a renormalizable field
  theory with unstable particles}},
  \href{https://doi.org/10.1016/S0031-8914(63)80277-3}{\emph{Physica}
  {\bfseries 29} (1963) 186}.

\bibitem{tHooft:1973wag}
G.~'t~Hooft and M.~J.~G. Veltman, \emph{{DIAGRAMMAR}}, {\emph{NATO Sci. Ser. B}
  {\bfseries 4} (1974) 177}.

\bibitem{Stockinger:2005gx}
D.~St{\"o}ckinger, \emph{{Regularization by dimensional reduction: consistency,
  quantum action principle, and supersymmetry}},
  \href{https://doi.org/10.1088/1126-6708/2005/03/076}{\emph{JHEP} {\bfseries
  0503} (2005) 076} [\href{https://arxiv.org/abs/hep-ph/0503129}{{\ttfamily
  hep-ph/0503129}}].

\bibitem{'tHooft:1972fi}
G.~'t~Hooft and M.~Veltman, \emph{{Regularization and Renormalization of Gauge
  Fields}},
  \href{https://doi.org/10.1016/0550-3213(72)90279-9}{\emph{Nucl.Phys.}
  {\bfseries B44} (1972) 189}.

\bibitem{Bollini:1972ui}
C.~G. Bollini and J.~J. Giambiagi, \emph{{Dimensional Renormalization: The
  Number of Dimensions as a Regularizing Parameter}},
  \href{https://doi.org/10.1007/BF02895558}{\emph{Nuovo Cim.} {\bfseries B12}
  (1972) 20}.

\bibitem{Collins:1984xc}
J.~C. Collins, \emph{Renormalization}. Cambridge University Press, 1984.

\bibitem{Bern:1991aq}
Z.~Bern and D.~A. Kosower, \emph{{The Computation of loop amplitudes in gauge
  theories}}, \href{https://doi.org/10.1016/0550-3213(92)90134-W}{\emph{Nucl.
  Phys.} {\bfseries B379} (1992) 451}.

\bibitem{Siegel:1979wq}
W.~Siegel, \emph{{Supersymmetric Dimensional Regularization via Dimensional
  Reduction}},
  \href{https://doi.org/10.1016/0370-2693(79)90282-X}{\emph{Phys.Lett.}
  {\bfseries B84} (1979) 193}.

\bibitem{Signer:2008va}
A.~Signer and D.~St{\"o}ckinger, \emph{{Using Dimensional Reduction for
  Hadronic Collisions}},
  \href{https://doi.org/10.1016/j.nuclphysb.2008.09.016}{\emph{Nucl.Phys.}
  {\bfseries B808} (2009) 88}
  [\href{https://arxiv.org/abs/0807.4424}{{\ttfamily 0807.4424}}].

\bibitem{Fazio:2014xea}
R.~A. Fazio, P.~Mastrolia, E.~Mirabella and W.~J. Torres~Bobadilla, \emph{{On
  the Four-Dimensional Formulation of Dimensionally Regulated Amplitudes}},
  \href{https://doi.org/10.1140/epjc/s10052-014-3197-4}{\emph{Eur. Phys. J.}
  {\bfseries C74} (2014) 3197}
  [\href{https://arxiv.org/abs/1404.4783}{{\ttfamily 1404.4783}}].

\bibitem{Battistel:1998sz}
O.~A. Battistel, A.~L. Mota and M.~C. Nemes, \emph{{Consistency conditions for
  4-D regularizations}},
  \href{https://doi.org/10.1142/S0217732398001686}{\emph{Mod. Phys. Lett.}
  {\bfseries A13} (1998) 1597}.

\bibitem{Cherchiglia:2010yd}
A.~L. Cherchiglia, M.~Sampaio and M.~C. Nemes, \emph{{Systematic Implementation
  of Implicit Regularization for Multi-Loop Feynman Diagrams}},
  \href{https://doi.org/10.1142/S0217751X11053419}{\emph{Int. J. Mod. Phys.}
  {\bfseries A26} (2011) 2591}
  [\href{https://arxiv.org/abs/1008.1377}{{\ttfamily 1008.1377}}].

\bibitem{Hernandez-Pinto:2015ysa}
R.~J. Hernandez-Pinto, G.~F.~R. Sborlini and G.~Rodrigo, \emph{{Towards gauge
  theories in four dimensions}},
  \href{https://doi.org/10.1007/JHEP02(2016)044}{\emph{JHEP} {\bfseries 02}
  (2016) 044} [\href{https://arxiv.org/abs/1506.04617}{{\ttfamily
  1506.04617}}].

\bibitem{Sborlini:2016gbr}
G.~F.~R. Sborlini, F.~Driencourt-Mangin, R.~Hernandez-Pinto and G.~Rodrigo,
  \emph{{Four-dimensional unsubtraction from the loop-tree duality}},
  \href{https://doi.org/10.1007/JHEP08(2016)160}{\emph{JHEP} {\bfseries 08}
  (2016) 160} [\href{https://arxiv.org/abs/1604.06699}{{\ttfamily
  1604.06699}}].

\bibitem{Sborlini:2016hat}
G.~F.~R. Sborlini, F.~Driencourt-Mangin and G.~Rodrigo, \emph{{Four-dimensional
  unsubtraction with massive particles}},
  \href{https://doi.org/10.1007/JHEP10(2016)162}{\emph{JHEP} {\bfseries 10}
  (2016) 162} [\href{https://arxiv.org/abs/1608.01584}{{\ttfamily
  1608.01584}}].

\bibitem{Pittau:2012zd}
R.~Pittau, \emph{{A four-dimensional approach to quantum field theories}},
  \href{https://doi.org/10.1007/JHEP11(2012)151}{\emph{JHEP} {\bfseries 11}
  (2012) 151} [\href{https://arxiv.org/abs/1208.5457}{{\ttfamily 1208.5457}}].

\bibitem{Donati:2013iya}
A.~M. Donati and R.~Pittau, \emph{{Gauge invariance at work in FDR: $H \to
  \gamma \gamma$}}, \href{https://doi.org/10.1007/JHEP04(2013)167}{\emph{JHEP}
  {\bfseries 04} (2013) 167} [\href{https://arxiv.org/abs/1302.5668}{{\ttfamily
  1302.5668}}].

\bibitem{Donati:2013voa}
A.~M. Donati and R.~Pittau, \emph{{FDR, an easier way to NNLO calculations: a
  two-loop case study}},
  \href{https://doi.org/10.1140/epjc/s10052-014-2864-9}{\emph{Eur. Phys. J.}
  {\bfseries C74} (2014) 2864}
  [\href{https://arxiv.org/abs/1311.3551}{{\ttfamily 1311.3551}}].

\bibitem{Pittau:2013qla}
R.~Pittau, \emph{{QCD corrections to $H \to gg$ in FDR}},
  \href{https://doi.org/10.1140/epjc/s10052-013-2686-1}{\emph{Eur. Phys. J.}
  {\bfseries C74} (2014) 2686}
  [\href{https://arxiv.org/abs/1307.0705}{{\ttfamily 1307.0705}}].

\bibitem{Page:2015zca}
B.~Page and R.~Pittau, \emph{{Two-loop off-shell QCD amplitudes in FDR}},
  \href{https://doi.org/10.1007/JHEP11(2015)183}{\emph{JHEP} {\bfseries 11}
  (2015) 183} [\href{https://arxiv.org/abs/1506.09093}{{\ttfamily
  1506.09093}}].

\bibitem{Gnendiger:2017pys}
C.~Gnendiger et~al., \emph{{To ${d}$, or not to ${d}$: recent developments and
  comparisons of regularization schemes}},
  \href{https://doi.org/10.1140/epjc/s10052-017-5023-2}{\emph{Eur. Phys. J.}
  {\bfseries C77} (2017) 471}
  [\href{https://arxiv.org/abs/1705.01827}{{\ttfamily 1705.01827}}].

\bibitem{Pittau:2013ica}
R.~Pittau, \emph{{On the predictivity of the non-renormalizable quantum field
  theories}}, \href{https://doi.org/10.1002/prop.201400079}{\emph{Fortsch.
  Phys.} {\bfseries 63} (2015) 132}
  [\href{https://arxiv.org/abs/1305.0419}{{\ttfamily 1305.0419}}].

\bibitem{Bednyakov:2016onn}
A.~V. Bednyakov, B.~A. Kniehl, A.~F. Pikelner and O.~L. Veretin, \emph{{On the
  $b$-quark running mass in QCD and the SM}},
  \href{https://doi.org/10.1016/j.nuclphysb.2017.01.004}{\emph{Nucl. Phys.}
  {\bfseries B916} (2017) 463}
  [\href{https://arxiv.org/abs/1612.00660}{{\ttfamily 1612.00660}}].

\bibitem{Gorishnii:1991zr}
S.~G. Gorishnii, A.~L. Kataev, S.~A. Larin and L.~R. Surguladze, \emph{{Scheme
  dependence of the next to next-to-leading QCD corrections to Gamma(tot) (H0
  $\to$ hadrons) and the spurious QCD infrared fixed point}},
  \href{https://doi.org/10.1103/PhysRevD.43.1633}{\emph{Phys. Rev.} {\bfseries
  D43} (1991) 1633}.

\bibitem{GehrmannDeRidder:2004tv}
A.~Gehrmann-De~Ridder, T.~Gehrmann and E.~W.~N. Glover, \emph{{Infrared
  structure of e$^+$e$^-$ $\to$ 2 jets at NNLO}},
  \href{https://doi.org/10.1016/j.nuclphysb.2004.05.017}{\emph{Nucl. Phys.}
  {\bfseries B691} (2004) 195}
  [\href{https://arxiv.org/abs/hep-ph/0403057}{{\ttfamily hep-ph/0403057}}].

\bibitem{Boughezal:2015ded}
R.~Boughezal, J.~M. Campbell, R.~K. Ellis, C.~Focke, W.~T. Giele, X.~Liu
  et~al., \emph{{Z-boson production in association with a jet at
  next-to-next-to-leading order in perturbative QCD}},
  \href{https://doi.org/10.1103/PhysRevLett.116.152001}{\emph{Phys. Rev. Lett.}
  {\bfseries 116} (2016) 152001}
  [\href{https://arxiv.org/abs/1512.01291}{{\ttfamily 1512.01291}}].

\bibitem{Magnea:2018ebr}
L.~Magnea, E.~Maina, G.~Pelliccioli, C.~Signorile-Signorile, P.~Torrielli and
  S.~Uccirati, \emph{{Factorisation and Subtraction beyond NLO}},
  \href{https://arxiv.org/abs/1809.05444}{{\ttfamily 1809.05444}}.

\bibitem{Vermaseren:1994je}
J.~A.~M. Vermaseren, \emph{{Axodraw}},
  \href{https://doi.org/10.1016/0010-4655(94)90034-5}{\emph{Comput. Phys.
  Commun.} {\bfseries 83} (1994) 45}.

\end{thebibliography}\endgroup
\bibliographystyle{JHEP}

\end{document}